\documentclass[usenatbib]{mnras}
\usepackage{graphicx,fixltx2e}
\usepackage{wasysym}
\usepackage{epsfig}
\usepackage{txfonts}
\usepackage{times}
\usepackage[T1]{fontenc}
\usepackage{ae,aecompl}
\usepackage{graphicx}	
\usepackage{color}
\usepackage{hyperref}

\def\Msun{\hbox{$M_{\odot}$}}             

\title[The magnetic properties of the planet host star Kepler-78]{The magnetic properties of the star Kepler-78\thanks{Based on observations obtained with ESPaDOnS at the Canada-France-Hawaii Telescope (CFHT) which is operated by the National Research Council (NRC) of Canada, the Institut National des Science de l'Univers of the Centre National de la Recherche Scientifique (CNRS) of France, and the University of Hawaii. }}

\author[]{
Moutou,C.$^{1,2}$\thanks{E-mail: moutou@cfht.hawaii.edu}, 
Donati J.-F.$^{3}$,
Lin D.$^{4}$,
Laine R.$^4$,
Hatzes A.$^5$\\
$^{1}$Canada France Hawaii Telescope Corporation, CNRS, 65-1238 Mamalahoa Hwy, Kamuela, HI 96743, USA\\
$^{2}$Aix Marseille Universit\'e, CNRS, LAM (Laboratoire d'Astrophysique de Marseille) UMR 7326, 13388, Marseille, France\\
$^{3}$Universit\'e de Toulouse / CNRS-INSU, IRAP / UMR 5277, Toulouse, F-31400 France\\
$^{4}$UC Santa Cruz, 1156 High Street, Santa Cruz, CA 95064, USA\\
$^5$ Th\"uringer Landessternwarte Tautenburg, Sternwarte 5, D-0, Germany}
\date{Accepted XXX. Received YYY; in original form ZZZ}
\pubyear{2016}

\begin{document}
\label{firstpage}
\pagerange{\pageref{firstpage}--\pageref{lastpage}}
\maketitle

\begin{abstract}

Kepler-78 is host to a transiting 8.5-hour orbit super-Earth. In this paper, the rotation and magnetic properties of the planet host star are studied. We first revisit the $Kepler$ photometric data for a detailed description of the rotation properties of Kepler-78, showing that the star seems to undergo a cycle in the spot pattern of $\sim$1,300 day duration. We then use spectropolarimetric observations with CFHT/ESPaDOnS to measure the circular polarization in the line profile of the star during its rotation cycle, as well as spectroscopic proxies of the chromospheric activity. The average field has a strength of 16 G. The magnetic topology is characterized by a poloidal and a toroidal component, encompassing 60\% and 40\% of the magnetic energy, respectively. Differential rotation is detected with an estimated rate of 0.105$\pm$0.039 rad.d$^{-1}$. Activity tracers vary with the rotation cycle of the star; there is no hint that a residual activity level is related to the planetary orbit at the precision of our data.  The description of the star magnetic field's characteristics then may serve as input for models of interactions between the star and its close-by planet, e.g., Ohmic dissipation and unipolar induction. 

\end{abstract}

\section{Introduction}
\label{intro}

Extrasolar planets at very short orbital distance from their stars have questioned the formation theories since their discovery 20 years ago \citep{mayorqueloz}, although a theoretical framework existed to explain their existence \citep{goldreich}: planets could be formed outside the snow line and migrate inwards due to interactions with the disk. After several decades of theoretical studies on planetary migration, these phenomena have been extensively described  \citep[for a recent review, see][ and references therein]{baruteaumasset}. Once started, the migration of the outer planets may occur on very short timescales. The efficiency of the mechanism thus requires a strong braking mechanism to halt the migration and insure the planet survival. It has been proposed that the migration stops when the planet enters the magnetospheric cavity of the star, where the disk is truncated \citep{lin96}. Making further progress in the description of these phenomena requires both to get measured constraints on this magnetospheric cavity and to probe the limit conditions of planet survival by discovering and characterizing extreme systems. \\

A new class of ultra-short-period planets has been recently unveiled by the $Kepler$ satellite: these are super-Earth candidates with orbital  periods of a few hours (106 candidates) \citep{sanchis14}. With such compact orbits, the evolution of these planets is strongly influenced by the radiative evaporation, tidal torque, and magnetic interaction with their host stars. The fact that the vast majority of these ultra-short-period candidates have radii smaller than 2 Earth radii could mean that they are either some failed cores or the residuals of more massive planets having suffered strong evaporation, mass loss, and/or tidal disruption. 

Following the recent theoretical developments on star-planet magnetic interactions, \citet{lainelin} have proposed that the motion of the small planet relative to the magnetic field of its host star may induce unipolar induction, similar to the mechanism between Io and Jupiter. This induction generates an electric field across the planet's surface and an electric current which circulates across the planet's mantle, along the flux tube between the planet and its host star, and across the footprint of the flux tube on the surface of the star. The intensity of the current and its associated Ohmic dissipation are determined by both the strength of the stellar magnetic field and the electric conductivity in the planetary mantle. In this study, we present new observations aimed at describing the magnetic properties and activity signatures of one host star of such planetary system, Kepler-78.

Kepler-78 is a 625$\pm$150 Myr, late G dwarf, which ultra-short-period planet has been detected and characterized in 2013 \citep{sanchis13,pepe,howard}. It is a planet very similar to Earth in radius, mass and density. Its extremely short orbital period of 8.5h period and 3R$_\star$ semi-major axis, however, makes it different to temperate telluric planets. The equilibrium temperature of the planet surface is of the order of 2300 to 3100K, depending  on its albedo, on its dayside \citep{sanchis13}. The rocks composing the telluric planet would be molten on the dayside, and condensed on the nightside, generating a desequilibrium of the planet structure and other extreme physical phenomena as first described by \citet{leger11} for the so-called "lava-ocean planets".

In this paper, we present an observational follow up of this interesting system. In Section 2, we revisit the $Kepler$ observations of this star. In Section 3, we present new observational data and analyses to constrain the magnetic properties of the planet host Kepler-78, obtained with ESPaDOnS at the Canada-France-Hawaii Telescope. In Section 4, we describe the results obtained by Zeeman Doppler Imaging on the magnetic topology of the stellar surface and differential rotation. In Section 5, we report the analysis of activity proxies from this data and search for potential planetary signatures. In Section 6, we present our search for reflected light signature. Then we summarize and present our conclusions in Section 7.

 \begin{table*}
\centering
\caption{Journal of the observations: UT date of observations, Julian date, rotational, orbital and synodic phases, total exposure time per visit, peak signal-to-noise ratio per 2.6 km/s pixel in the combined spectra, activity measurement proxies (CaII H and K, H$\alpha$ and CaII IRT), radial velocities and status of magnetic detections. Rotation period is 12.588 days, orbital period is 0.355074 day and synodic period is 0.365 day. Radial velocities (RV) have typical errors of 30m/s. In the last column, D stands for Definite detection (defined by a false alarm probability smaller than 10$^{-5}$), M for Marginal detection (false alarm probability between 10$^{-5}$ and 10$^{-3}$) and N for Non-detection of polarized signatures.}
\begin{tabular}{lccccccccccc}
\hline
UT Date & HJD-2400000. & $\Phi_{rot}$ & $\Phi_{orb}$ & $\Phi_{syn}$ & Texp &   SNR & H$\alpha$& CaII HK    & CaII IRT   & RV & Magn\\
          &days                   &                        &                        &                          & sec    &            &         &                      &   & km/s & Det \\
\hline
07Sep2014 &56907.889     &  -25.34  &  0.91 & 0.68   &4$\times$1700&293& & &&-3.477&N\\
12Sep2014 &56912.861     &  -24.95  &14.91 & 14.29&4$\times$1700&253& & & &-3.474&M\\
\hline
23Jul2015   &57226.905     & 0.000& 0.53   & 0.96      &4$\times$1715&275&-0.063 &+0.362	& +0.071&-3.429&M\\
		   &			    &		&		&		&			  &	  &$\pm$0.017&$\pm$0.087&$\pm$0.009& & \\
24Jul2015   &57227.893     & 0.079& 3.31   &  3.67     &4$\times$1715&280&-0.016&+0.620& +0.097&-3.426&M\\
		   &			    &		&		&		&			  &	  &$\pm$0.014&$\pm$0.067&$\pm$0.008& & \\
25Jul2015   &57228.979     & 0.165& 6.37   &  6.64     &4$\times$1715&284&+0.142&+0.051 & +0.075&-3.420&M\\
		   &			    &		&		&		&			  &	  &$\pm$0.014&$\pm$0.051&$\pm$0.013& & \\
26Jul2015   &57229.984     & 0.245& 9.20   &  9.39     &4$\times$1715&279&+0.000&+0.374& +0.050&-3.426&D\\
		   &			    &		&		&		&			  &	  &$\pm$0.015&$\pm$0.057&$\pm$0.011& & \\
27Jul2015   &57230.907     & 0.318&11.80  &  11.92   &4$\times$1715&276&-0.170&-1.385& -0.072&-3.439&D\\
		   &			    &		&		&		&			  &	  &$\pm$0.014&$\pm$0.110&$\pm$0.007& & \\
28Jul2015   &57231.977     & 0.403&14.81  &  14.85   &4$\times$1715&258&-0.213&-0.847& -0.096&-3.441&D\\
		   &			    &		&		&		&			  &	  &$\pm$0.015&$\pm$0.061&$\pm$0.010& & \\
29Jul2015   &57233.037     & 0.487&17.80  &  17.75   &4$\times$1715&274&-0.464&-0.452&-0.152&-3.428&M\\
		   &			    &		&		&		&			  &	  &$\pm$0.022&$\pm$0.054&$\pm$0.015& & \\
30Jul2015   &57233.863     & 0.553&20.13  &  20.01   &4$\times$1715&276&-0.323&-0.529& -0.062&-3.446&M\\
		   &			    &		&		&		&			  &	  &$\pm$0.015&$\pm$0.059&$\pm$0.008& & \\
31Jul2015   &57234.824     & 0.629&22.83  &  22.64   &4$\times$1715&267&-0.078&-0.359& -0.055&-3.439&N\\
		   &			    &		&		&		&			  &	  &$\pm$0.014&$\pm$0.061&$\pm$0.009& & \\
01Aug2015 &57235.895     & 0.714&25.85  &  25.57   &4$\times$1715&271&-0.022&-0.272& -0.110&-3.448&N\\
		   &			    &		&		&		&			  &	  &$\pm$0.017&$\pm$0.068&$\pm$0.010& & \\
27Aug2015 &57261.836     & 2.775&98.92  &  96.58   &4$\times$1715&273&+0.486&+0.405& +0.074&-3.461&D\\
		   &			    &		&		&		&			  &	  &$\pm$0.022&$\pm$0.049&$\pm$0.012& & \\
28Aug2015 &57262.800     & 2.852&101.64&  99.22   &4$\times$1715&286&+0.523&+0.773& +0.095&-3.454&M\\
		   &			    &		&		&		&			  &	  &$\pm$0.021&$\pm$0.069&$\pm$0.013& & \\
29Aug2015 &57263.882     & 2.938&104.68&102.18   &4$\times$1715&276&+0.181&-0.068& +0.049&-3.452&N\\
		   &			    &		&		&		&			  &	  &$\pm$0.021&$\pm$0.067&$\pm$0.011& & \\
\hline
\end{tabular}
\label{log}
\end{table*}

\section{Photometric variability from $Kepler$ observations}
Kepler-78 (also known as TYC 3147-188-1 or KIC 8435766) was observed by the $Kepler$ satellite for almost 4 years. Due to an early limitation of the $Kepler$ period-search space towards short periods, the planet signal was not primarily part of the list of $Kepler$ Objects of interest  \citep{batalha}. Further studies, however, focused on the search for transits at extremely short orbital periods \citep[e.g.,][]{sanchis14}, and their first success was the discovery of Kepler-78 b \citep{sanchis13}. Both the primary transit, the occultation and the phase variations during the 8.5-hour orbit were detected. Kepler-78 b mass was then measured after intensive radial velocity campaigns were carried out using HARPS-N on the Telescope National Galileo \citep{pepe} and HIRES on the Keck telescope \citep{howard}, although stellar activity has been a serious limitation of these analyzes. Both spectroscopic studies, using different data sets and independent methods for the correction of the stellar jitter, resulted in very similar parameters for the planet mass (1.86$\pm$0.25 and 1.69$\pm$0.41 Earth mass, respectively, in circular orbits). \citet{grunblatt} and \citet{hatzes} also determined planet masses from the same data set. Combined with the radius inferred by $Kepler$, the planet density is very similar to Earth's density, although Kepler-78 b evidently has had a very different orbital evolution and most probably has a different internal structure.\\

Concerning the stellar properties, Kepler-78 has been identified as a young, late G dwarf, with an age of about 625$\pm$150 Myr and a stellar mass of 0.76-0.83 \Msun\ \citep{sanchis13,pepe,howard}. Due to a relatively young age, the star is active as seen from its $Kepler$ lightcurve. The mean rotation period of the star based on the lightcurve autocorrelation analysis has been found to be 12.588$\pm$0.03 days \citep{mcquillan14}. Despite the low error bar associated with this measurement, it is clear from the lightcurve that a single period cannot explain all the observed variability.\\

\begin{figure*}
\centering
\includegraphics[width=0.45\columnwidth]{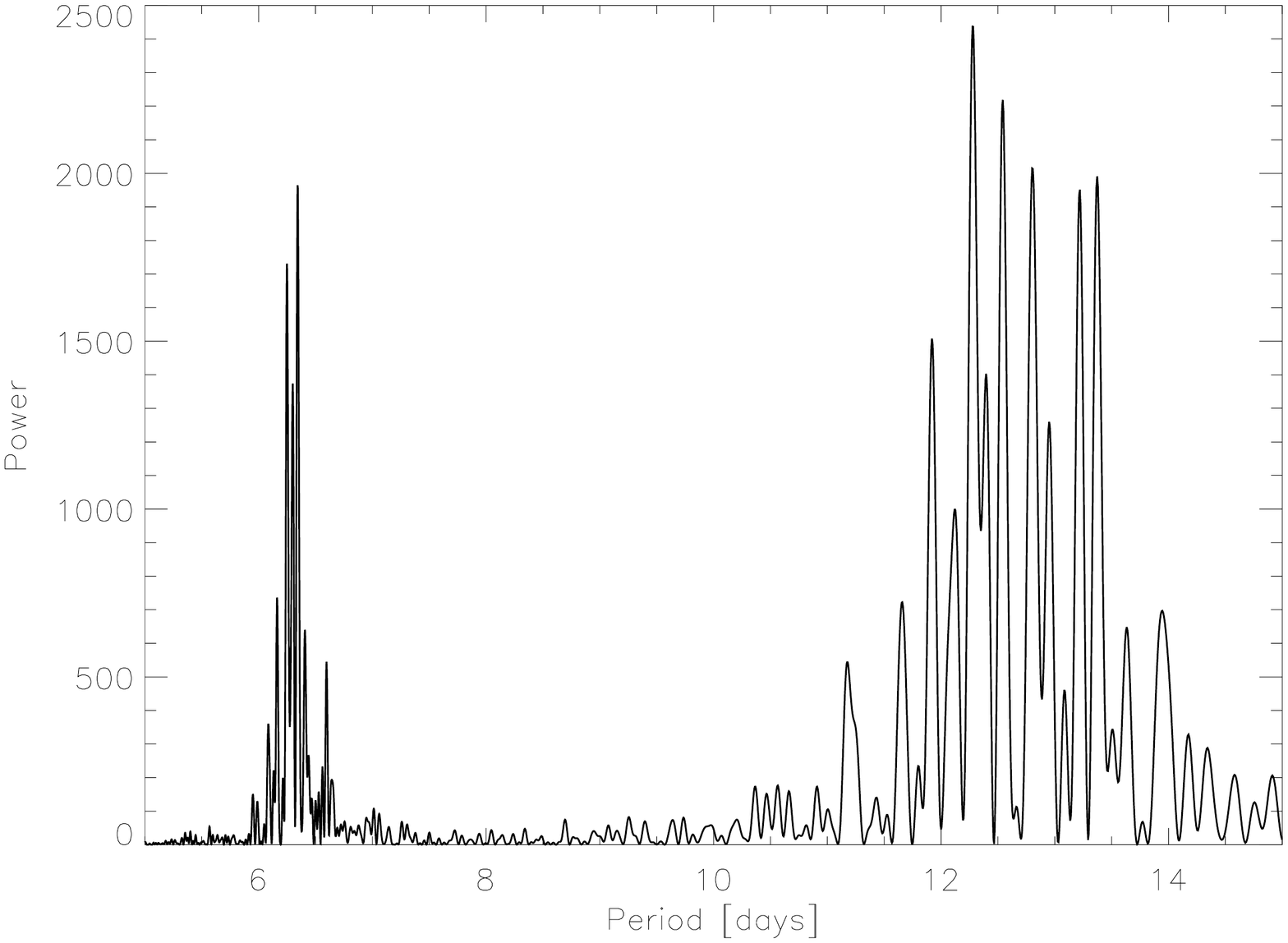}
\includegraphics[width=0.45\columnwidth]{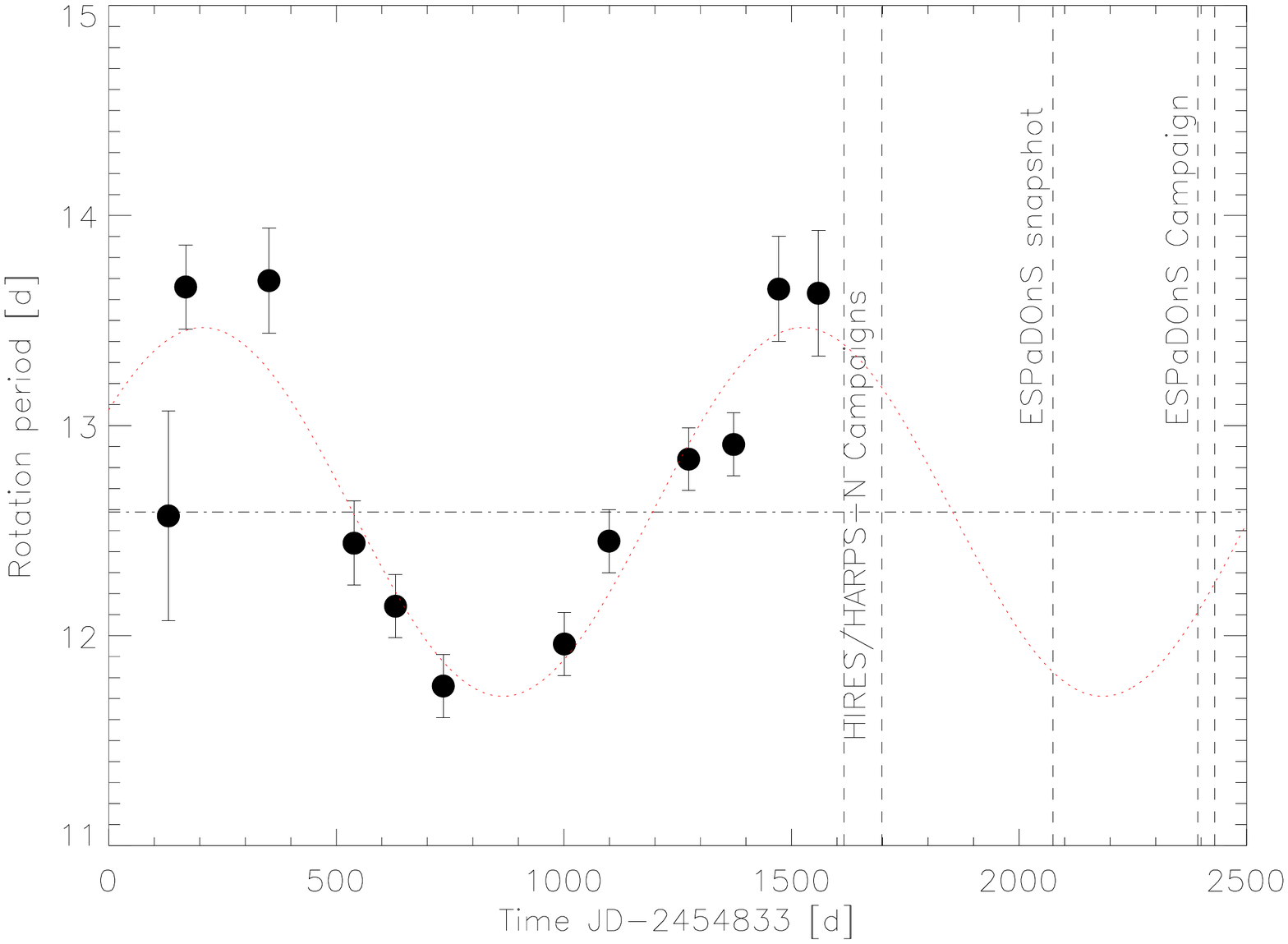}
\caption{(Left) The periodogram of the full $Kepler$ lightcurve of Kepler-78, zoomed around the rotational period and its first harmonic. (Right) The modulation in period of the main peak of the periodogram, measured on each quarter individually, as a function of time. The time of the subsequent spectroscopic campaigns is also noted. The sinusoidal dotted line is a  fit of these values, that may be interpreted as evidence for an activity cycle (see text) while the horizontal dot-dash line shows the mean rotation period found by \citet{mcquillan14}. }
\label{rot}
\end{figure*}

We have computed the periodograms of $Kepler$ data using the Exoplanet Archive online tool\footnote{http://exoplanetarchive.ipac.caltech.edu/cgi-bin/Pgram/nph-pgram} for the 4 years of data (Figure \ref{rot}, left). The main peak covers the range 11 to 14.5 days, and a secondary set of peaks at about half the main period, between 6 and 6.5 days.  Then we did the computation quarter by quarter, excluding the first slot of data which covers less than a stellar period (see Figure \ref{appenfig1} for each individual periodogram). Figure \ref{rot} (right) shows the location of the main peak of these periodograms as a function of time, compared to the mean rotation period of 12.588$\pm$0.03 days. Error bars were also estimated from the periodograms. For some quarters, the main peak is not at the full period, but one of the harmonics (not shown on the figure, see Figure \ref{appenfig1} ). When the main peak shows the rotation period, this value seems to be first decreasing,  then increasing, showing a maximum of 13.8 days at about JD=245433+250 and +1550. The minimum is around 11.7 days. We don't include the quarters where the main periodogram peak corresponds to about half the rotation period (doubling the period), because then the stellar surface probably has two main active regions which do not have to lie exactly 180$^\circ$ apart, so twice these periods do not correspond exactly to the full rotational period -- or the $\sim$12 day period would show a significant power in the periodogram. \\

Apart from one deviant point at the beginning of the series (which corresponds to a shorter period of time, 33 days instead of a full quarter of 90 days, hence a large error bar), the behaviour of the measured rotational periods looks sinusoidal. We may interpret this behaviour as a stellar activity cycle, similar to the solar cycle. Due to some amount of differential rotation and to the dynamo cycle of the star, the spots appearing at higher latitude at the beginning of the $Kepler$ time series have a 13.8-day period, then migrate to the equator where their rotation period is shorter (down to 11.7 days). When fitting a sine wave, we find a period of 1300$\pm$200 days, amplitude of 0.9$\pm$0.2 days and a mean period of 12.57$\pm$0.17 days. 
If what we see on Kepler-78 is similar in nature to the solar cycle, we expect maximum activity to correspond to epochs where the apparent rotation period is smallest, i.e. when spots tend to cluster near the equator. 

If our interpretation is correct, we could be witnessing an activity cycle of about 3.5 year duration, three times shorter than the 11-yr solar cycle. A sophisticated spot modeling of the full lightcurve could probably reproduce the butterfly diagram of Kepler-78, which is not the purpose of this paper. Two values could be derived from this analysis: i) the differential rotation is at least 0.08 rad.d$^{-1}$ (since spots probably cover only a range of latitude and do not extend all the way to the poles, it is a lower limit), and ii) the mean rotational period of the star during our ESPaDOnS campaign can be extrapolated to 12.2$\pm$0.3 days (shown by the vertical lines on the right of Figure \ref{rot}, right, assuming the period and amplitude of the cycle are as quoted above). We could also expect most of the active regions (or the dominating ones) to be near the $\sim$30$^\circ$ latitude at this epoch (assuming a differential rotation law in sin$^2$$\theta$ where $\theta$ is the latitude of the region, see Section 4.1).
The rotation period during the HARPS-N and HIRES campaigns may have been close to the maximum rotation period ($\sim$13.5 days) in our extrapolation, and probably close to a minimum of the stellar spot activity. 
In his pre-whitening analysis, \citet{hatzes} finds maximum power at 6.44 days in HIRES data and 3.42 days in HARPS-N data. These frequencies are interpreted as P$_{rot}$/2 and P$_{rot}$/4, respectively, which points towards a rotational period in the range 12.7 to 13.6 days. The detection of the rotational period in the radial velocity data is, however, not as precise as with $Kepler$, because of the sparse sampling of these two data sets.

\section{Spectropolarimetric observations}
\label{espadons}
We observed Kepler-78 with the ESPaDOnS spectropolarimeter  at the Canada-France-Hawaii Telescope ontop of Mauna Kea. 
In its polarimetric mode, ESPaDOnS allows us to observe the spectral range from 370 to 1050 nm at a resolving power of 65,000 (see Figure \ref{appenfig3} for an example spectrum). We measured the circularly-polarized signal in the Stokes $V$ configuration as well as the Stokes $I$ unpolarized spectra. Obtaining the circular polarization is achieved by acquiring sequences of four consecutive exposures in different polarimeter configurations which allows not only to detect polarisation signatures for the chosen Stokes parameter, but also to determine the potential level of spurious signatures  in the null profiles N (and correct for them if needed). 
 
Two sequences were first acquired in September 2014\footnote{Discretionary Time Program 14BD91}, during which feasibility was assessed. Since the polarized signal was detected in these spectra, we applied for more telescope time to observe Kepler-78 with ESPaDOnS over a full rotational cycle (about two weeks). These observations were performed between July 22 and August 28, 2015\footnote{Using the programs 15AC21, 15AF09, and 15BD96}. We secured 13 spectropolarimetric sequences with a peak signal-to-noise ratio ranging from 250 to 285 per 2.6 km/s pixel, at 780nm. The last three exposures are separated from the first ten by about a month. 
Since the last three sequences nicely fill the gap in the phase coverage of the rotational cycle, in the following, we analyze the data set of the all 13 sequences acquired in 2015. The log of the observations is in Table \ref{log}.
Reference ephemeris for the phases are $T_0$ = 2,457,226.9, P$_{rot}$ = 12.588 days \citep{mcquillan14} for the stellar rotation and P$_{orb}$ = 0.355074 day and T$_{0}$ = 2,454,953.95995 (transit time) for the planet orbit \citep{sanchis13}.\\
 
The ESPaDOnS data were processed by the  Libre-Esprit pipeline \citep{donati97,donati07}, which includes flatfield correction, wavelength calibration and optimal extraction of all (highly distorted) orders along the (tilted and wiggly) slit formed by the image slicer at the entrance of the spectrograph. The pipeline also provides a correction for the instrumental spectral drift using the telluric spectrum achieving an average $rms$ radial-velocity precision of 20-30m/s \citep{moutou07}. We then used the Least Square Deconvolution technique \citep{donati97} to take advantage of the very large number of stellar lines in the spectrum (6740 in average). We used a mask of stellar lines from Atlas9 LTE model atmosphere \citep{kurucz93} featuring a 5000K star with $log$g=4.5. This provides us with a mean intensity profile for each sequence, where the radial velocity of the star can be measured (Table \ref{log}). All spectra are corrected for spectral shifts resulting from instrumental effects using telluric lines as a reference. The $rms$ of the radial velocities is 12 m/s and the peak-to-peak variation is 40 m/s, while individual measurement accuracies range from 30 to 40m/s. Kepler-78 radial velocity jitter in previous studies is found to be characterized by a standard deviation of 4.3 and 10 m/s, respectively, for \citet{pepe} and \citet{howard}. Peak-to-peak variations of 23 and 49 m/s were found in these data sets. Systemic velocity is found to be -3.44 km/s in our data, and -3.51 km/s (respectively, -3.59 km/s) for these previous studies. The difference in systemic velocity are mostly due to instrumental shifts, since different instruments and pipelines are used (and ESPaDOnS has no absolute velocity calibration), and to activity. Our values of RV jitter  and peak-to-peak variations are similar to those measured in previous data sets. This is not the goal of the paper to try to correct for the jitter and search for the planet-induced Doppler signal. This is not possible given the limited RV precision of ESPaDOnS and the short time series. 

Polarized signatures in Stokes V profiles are corrected for small residual signals in null profiles N, such corrections having no more than a small impact on final 
results.  Zeeman signatures are detected in 4 of Stokes V Least Square Deconvolution profiles, featuring amplitudes of less than 0.05\% of the continuum. 
There are marginal detections in 6 others and no detection in 3 profiles. Marginal and definite detections correspond to a false-alarm probability threshold of, respectively, 10$^{-3}$ and 10$^{-5}$.
We show in Fig. \ref{stokesv} our sets of Stokes V Least Square Deconvolution profiles.


\section{Magnetic imaging and differential rotation }
\subsection{Method}
To reconstruct the magnetic maps of Kepler-78, we use the tomographic Zeeman Doppler Imaging (ZDI) code. It is set up to invert the time series of Stokes V profiles into magnetic maps of the stellar surface, i.e. the distribution of magnetic fluxes and orientations \citep{donati97}.  The magnetic map is iteratively derived from a comparison with the observed and modeled Stokes V profiles until we reach a pre-defined $\chi^2_r$ level. The ZDI code decomposes the stellar surface into multiple grid cells of similar projected areas and calculates the contribution of each grid to the spectral profile. The reconstructed profiles are obtained by summing the spectral contribution of all cells, for each rotation phase. Given that the problem is ill-posed, we use the maximum-entropy criterion to select the image with the least amount of information among all those fitting the data at the required level. Like Doppler Imaging, ZDI relies on the assumption that the observed variability is solely attributable to rotational modulation, and potentially to differential rotation (if included in the modeling);  any source of intrinsic variability beyond differential rotation cannot be explained with ZDI. The ZDI code has been based on early work by \citet{skilling84}, \citet{brown91} and \citet{donati97} and further developed in the last twenty years from magnetic studies of stars of various properties \citep[e.g.][]{donati06,donati08,morin10,fares10}. We refer the reader to these studies for further details on the method.

\begin{figure}
\centering
\includegraphics[width=0.5\columnwidth]{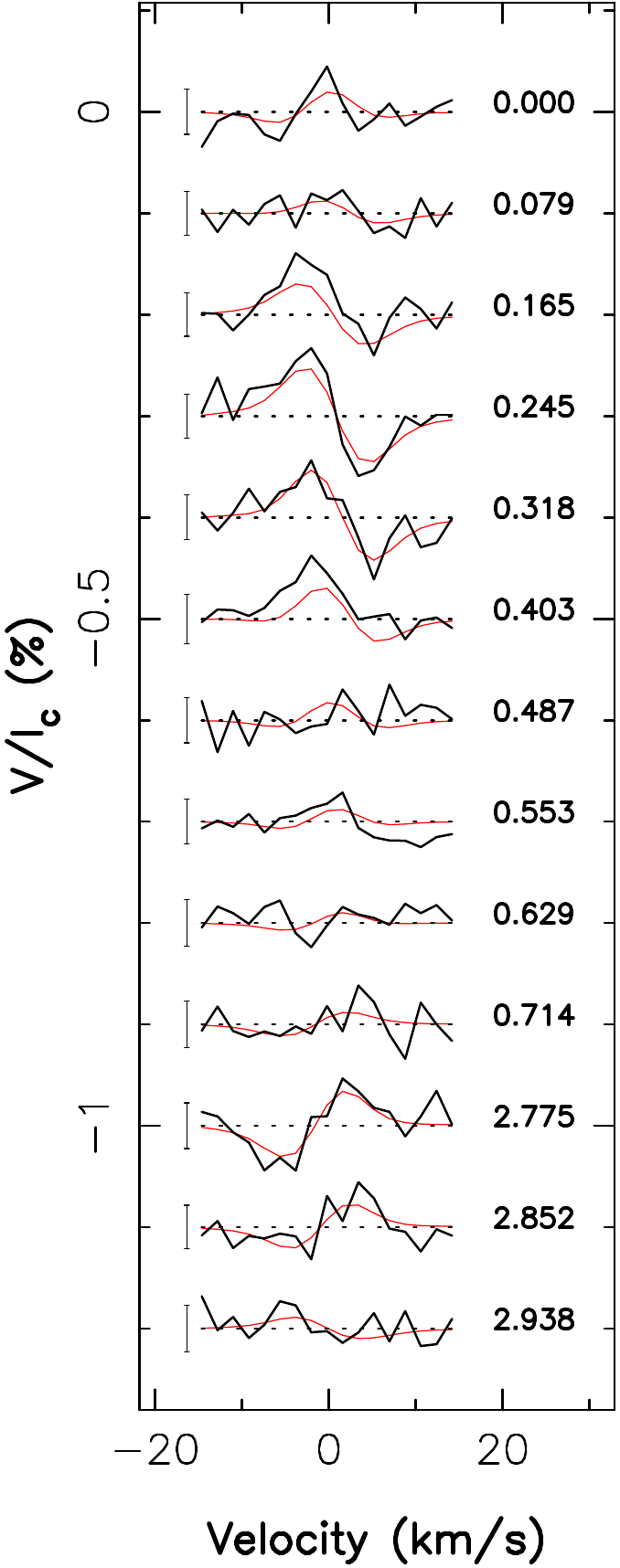}
\caption{ Circular polarization profiles of Kepler-78. The observed and synthetic profiles are shown in black and red respectively. On the left of each profile we show a $\pm$1$\sigma$ error bar, while on the right the rotational cycles are indicated.}
\label{stokesv}
\end{figure}

 In the ZDI code, the magnetic field is described by its radial poloidal, non-radial poloidal and toroidal components, all expressed in terms of spherical harmonics expansion. In the case of Kepler-78, we truncated the spherical harmonics expansion to the 5 lowest terms ($\ell$<5), which is adequate for slow rotators \citep[e.g.][]{fares10}. 
Fig. \ref{stokesv} shows our fit of the data using the ZDI modeling. We adopted the $v$sin$i$  value of 3$\pm$1 km/s, from literature values \citep{howard}, with little impact on the ZDI analysis. For the inclination, we used i=80$^\circ$, assuming that the orbit of the planet is more or less aligned with the rotation of the star.  The star inclination is known to have little impact on the output of the reconstruction, within typically 10-15$^\circ$.
The fits we obtain correspond to a $\chi^2_r$ equal to 1.0 starting for a $\chi^2_r$  of 2.1.  
A map in brightness cannot be obtained for Kepler-78 which has a too small rotation rate and too low profile distortions. 

The fit to the data is slightly improved if we assume that the star is rotating differentially rather than as a solid body (improvement of 1.9\% of the $\chi^2_r$ compared to the configuration where rotation is solid with a 12.588d period). We proceed as in our previous studies \citep[e.g.][]{donati00,petit02,donati03,donati14}. For a differentially rotating star, magnetic regions located at different latitudes have different angular velocities. In the model, the rotation rate at the surface of the star may vary with latitude $\theta$ as in the Sun, i.e., $\Omega$($\theta$) = $\Omega_{eq}$ -- d$\Omega$ sin$^2$$\theta$, where $\Omega_{eq}$ is the rotation rate at the equator  and d$\Omega$ the difference in rotation rate between the equator and the pole. We may thus estimate the spectral contribution of each elementary region at the surface of the star to the synthetic disk-integrated Stokes V stellar profiles for given values of $\Omega_{eq}$ and d$\Omega$. For each pair of ($\Omega_{eq}$, d$\Omega$), we then reconstructed a magnetic image at a given information content from the observed profiles and get the $\chi^2_r$. The obtained surface of $\chi^2_r$ is fitted by a paraboloid to obtain the optimum rotation values of the star (Fig. \ref{dr}).

\subsection{Results}
\label{reszdi}
\begin{figure}
\centering
\includegraphics[width=0.75\columnwidth]{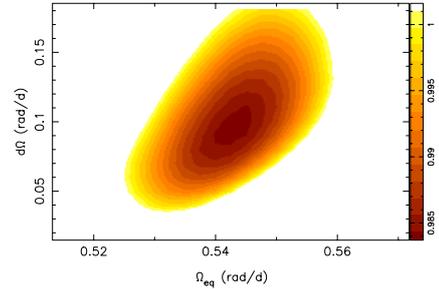}
\caption{ Variations of $\chi^2_r$ as a function of $\Omega_{eq}$ and d$\Omega$, derived from the
modelling of our observed profiles. A well-defined paraboloid is observed, with the outer colour contour
tracing the 2$\sigma$ ellipse for both parameters as a pair.}
\label{dr}
\end{figure}

The optimal rotation parameters derived from the magnetic reconstruction as shown in Fig. \ref{dr} are $\Omega_{eq}$ = 0.543$\pm$0.009 rad.d$^{-1}$ and d$\Omega$ = 0.105$\pm$0.039 rad.d$^{-1}$. There is thus a hint of detection of  differential rotation in the data set, at the 2$\sigma$ level. More specifically, the imaging code is found to have little leverage to reconcile all observed Zeeman signatures (e.g., the null and negative signals at cycles 0.714 and 2.775) with a consistent large-scale parent magnetic topology without invoking some amount of differential rotation and / or some level of intrinsic variability at the surface of the star.  The derived differential rotation parameters (see Fig \ref{dr}) correspond to an equator rotational period of 11.6$\pm$0.2 days and a pole rotation period of 14.3$\pm$1.3 days. These values are in good agreement with the ones obtained from the $Kepler$ lightcurve (section 2), 11.7 and 13.8$\pm$0.3 days for the minimum and maximum apparent rotation periods seen in the spot patterns along the 4-yr time span, giving us further confidence that what was measured with ZDI is indeed mostly differential rotation rather than intrinsic variability. The spots responsible for the photometric variations thus have a mean latitude varying from $\sim$15$^\circ$ to  $\sim$65$^\circ$. On the Sun, as a comparison, the differential rotation d$\Omega$ is 0.055 rad.d$^{-1}$ and the range of latitudes where the main spots emerge is from 5 to 30$^\circ$ \citep{hathaway2011}.  So, Kepler-78 has a twice stronger rotation shear and its spots expand over a wider range of latitudes.

\begin{figure*}
\centering
\includegraphics[width=\textwidth]{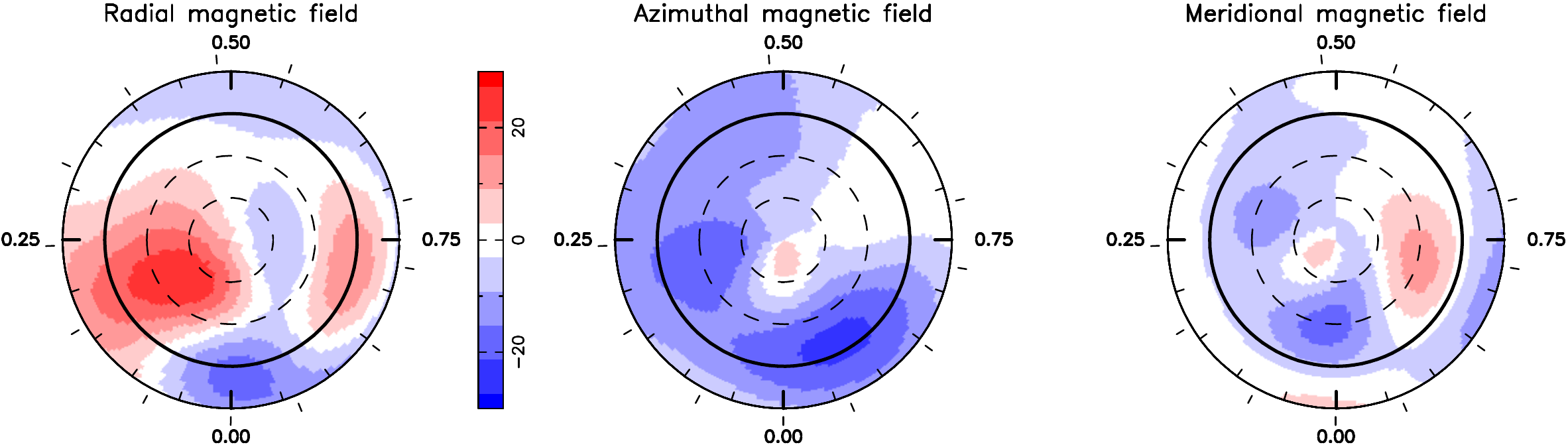}
\caption{Magnetic map of Kepler-78. The three components of the field in spherical coordinates system in a flattened polar view of the star are presented, down to latitude -30$^\circ$. The bold circle represents the equator. The small radial ticks around the star represent the rotational phases of our observations. The radial, azimuthal and meridional field maps are labeled in G and have the same color scale. }
\label{map}
\end{figure*}

The reconstructed magnetic map including the differential rotation is shown in Fig. \ref{map}. 
The main component of the map is a wide feature spanning from the equator up to 60$^\circ$ latitude visible in the radial field at the rotational phase 0.18.
The magnetic map features a dipole field having an amplitude of 12 G tilted at 50$^\circ$ to the rotation axis, as well as quadrupole and octupole components being both 
comparable in strength with the dipole (12 G ad 10 G, respectively). The average field strength is 16 G. The field has a predominant poloidal component while the fractional energy in the toroidal field amounts to $\sim$40\% of the total magnetic energy. The poloidal field component is mainly non-axisymetric, with $\sim$60\% of its magnetic energy in modes with $m > \ell/2$.\\
 
The magnetic topology of Kepler-78 is found to be typical for a 0.8\Msun\ star with this rotation period, in agreement with the diagram shown in Figure 3 of \citet{donatilandstreet}.  In particular, the
significant toroidal field and non-axisymmetric poloidal field are properties shared by such types of stars. 
We may compare to the sample presented in \citet{fares13}, where stars most similar to Kepler-78 are HD 102195 (mean field strength of 12G, 44\% of magnetic energy in the poloidal component and 25\% axisymmetry in the poloidal field) and HD 189733 observed at different epochs (22 to 36G, 33 to 77\% and 17 to 56\%, respectively). With values of [16G, 60\% and 60\%], the field of Kepler-78 is comparable in strength, and has more axisymmetry in its poloidal component. As seen with HD 189733 though, these values evolve on the year timescales for such stars and may span a wide range \citep{fares10}. 

\begin{figure}
\centering
\includegraphics[width=0.75\columnwidth,angle=270]{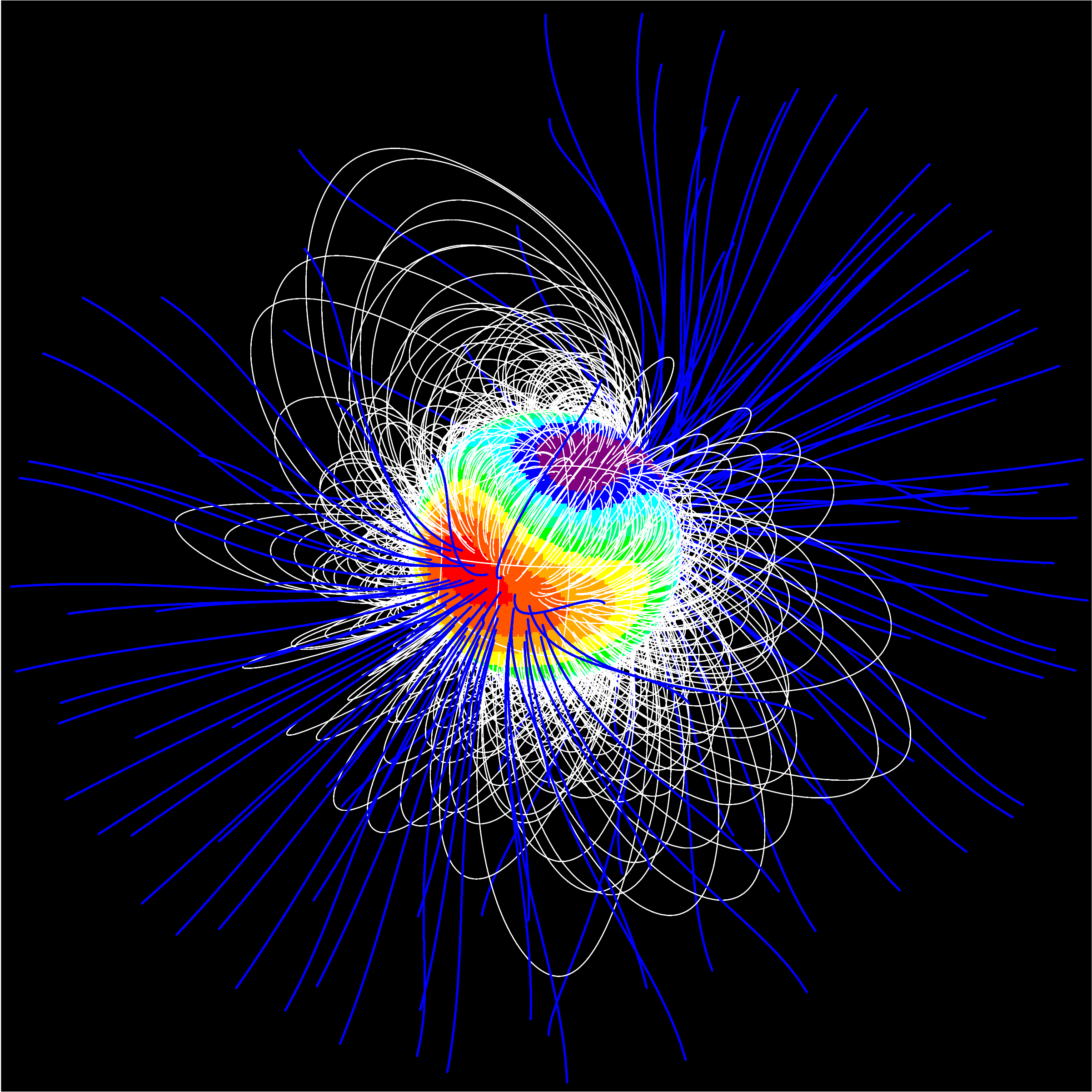}
\caption{ The extrapolated magnetic field of Kepler-78. White lines corresponds to the closed magnetic lines and blue ones to the open field lines. The star is shown as viewed from the observer at rotation phase 0.1.  An animated image is available at \url{http://www.ast.obs-mip.fr/users/donati/kepler78_jul15_dr.gif}}
\label{extrapolation}
\end{figure}

From the surface magnetic field, one can then extrapolate the magnetic  field  in  the  stellar  atmosphere  assuming  a  potential  field \citep{jardine2002}. This assumption is mildly relevant given the contribution of the toroidal field. In this technique, we assume that there is a source surface beyond which the field becomes purely radial; closed field lines are inside this volume. In this analysis, the source surface is put at a distance of 4 stellar radii from Kepler-78, which includes the planet orbit (at 3 stellar radii). The planet thus crosses regions where the field lines are open and others where they are closed. Figure \ref{extrapolation} shows the extrapolation features obtained with these assumptions, as would be seen for the rotational phase 0.1. The magnetic field strength seen by the planet strongly depends on the position of this reference source surface with respect to the planet orbit. While we cannot easily predict the amplitude of magnetic variations seen by Kepler-78 b, we know at which phases it sees a null energy, modulo some uncertainties due to the stellar inclination. Both the location of the source surface and the ratio of the dipole/quadrupole/octupole components are critical to characterize the field topology at the planet location. The latter ratio is not determined very accurately from the present data set given the temporal coverage.  


Relative motion between the planet and the time-averaged stellar field leads to unipolar induction. 
Permeation of time-dependent stellar field into the planet's interior also leads to Ohmic dissipation.
Both effects can induce heating of the planet's interior and orbital decay over few Myrs due to the Lorentz torque \citep{lainelin08,lainelin}.

\section{Activity proxies}

\begin{figure*}
\centering
\includegraphics[width=0.45\columnwidth]{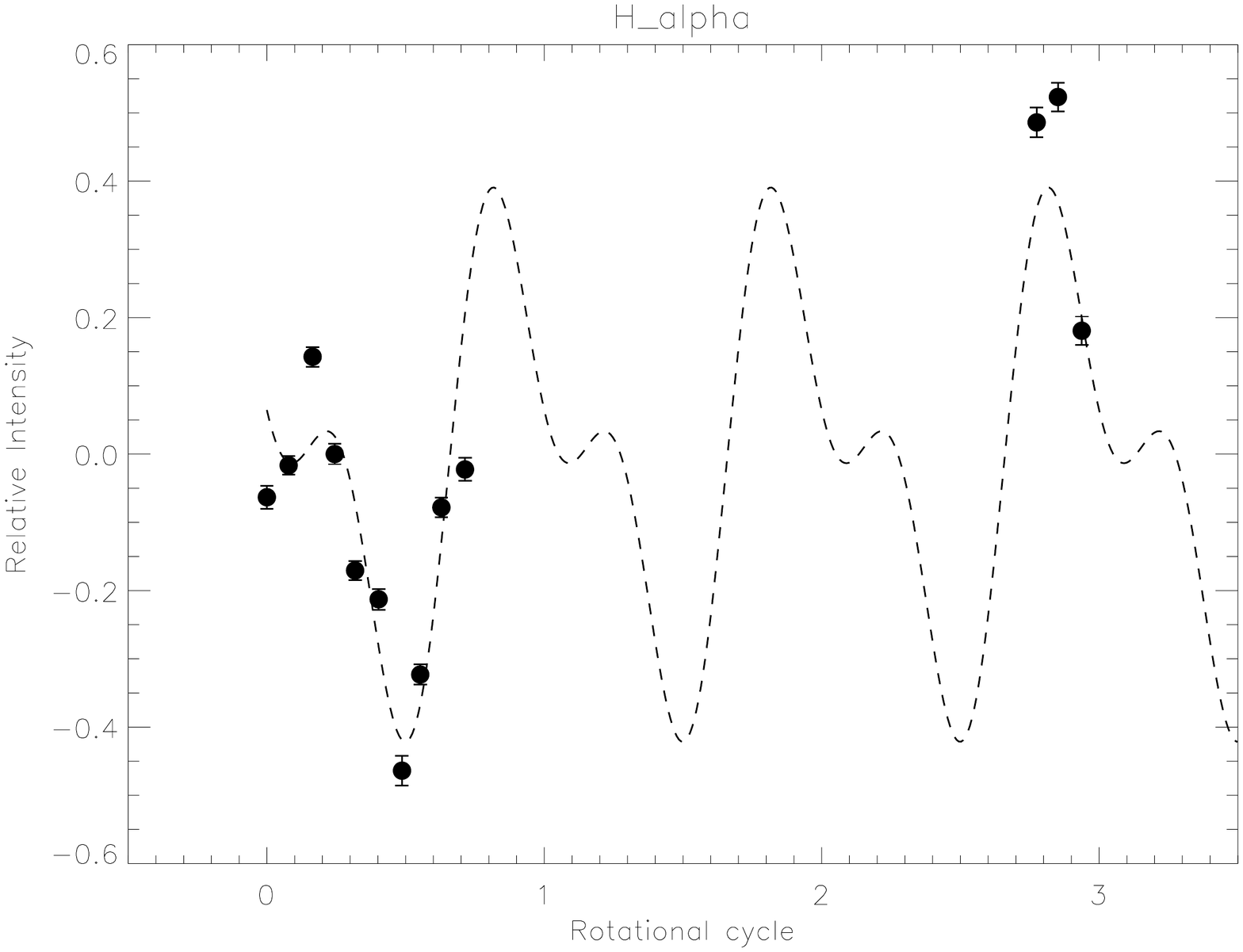}
\includegraphics[width=0.45\columnwidth]{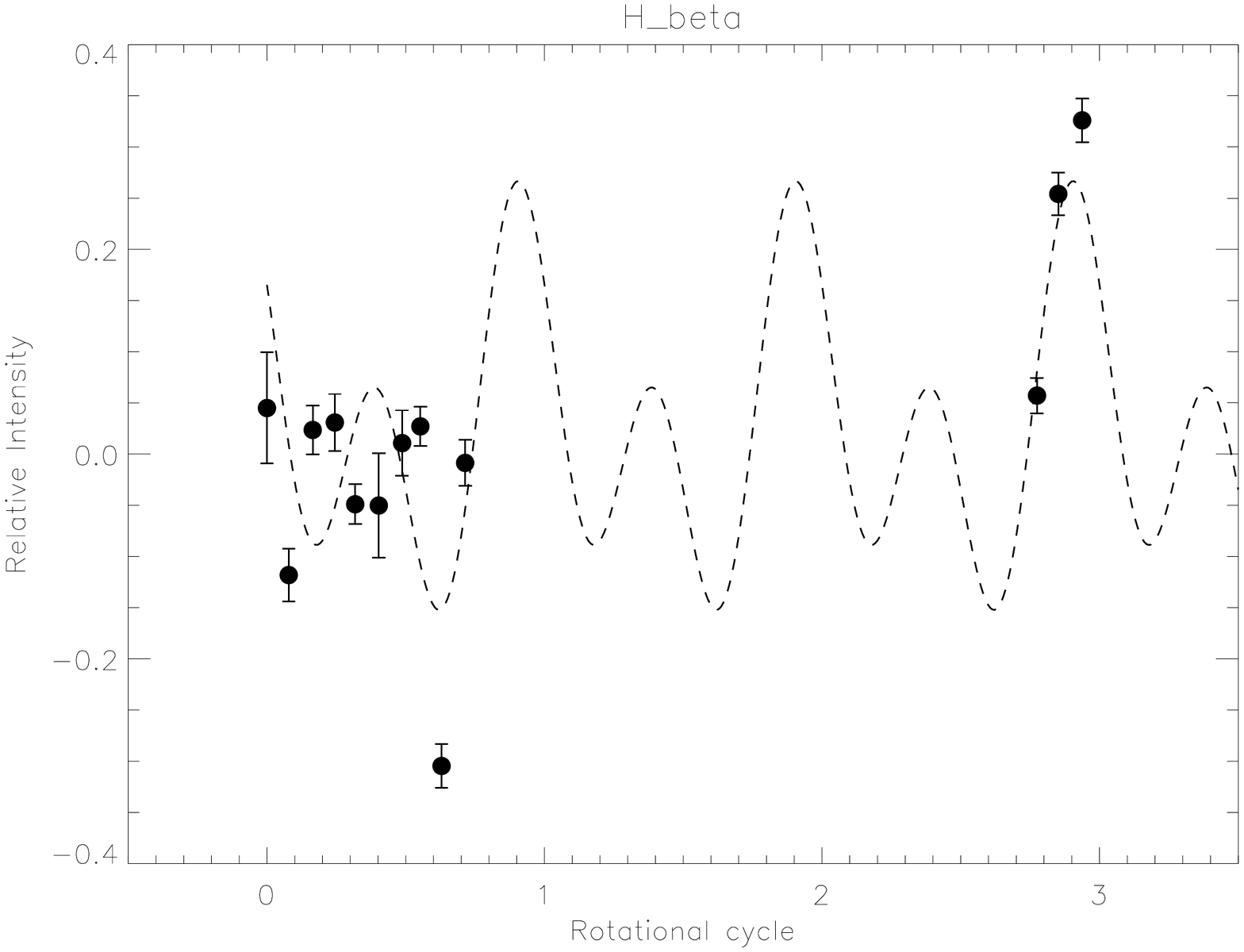}
\includegraphics[width=0.45\columnwidth]{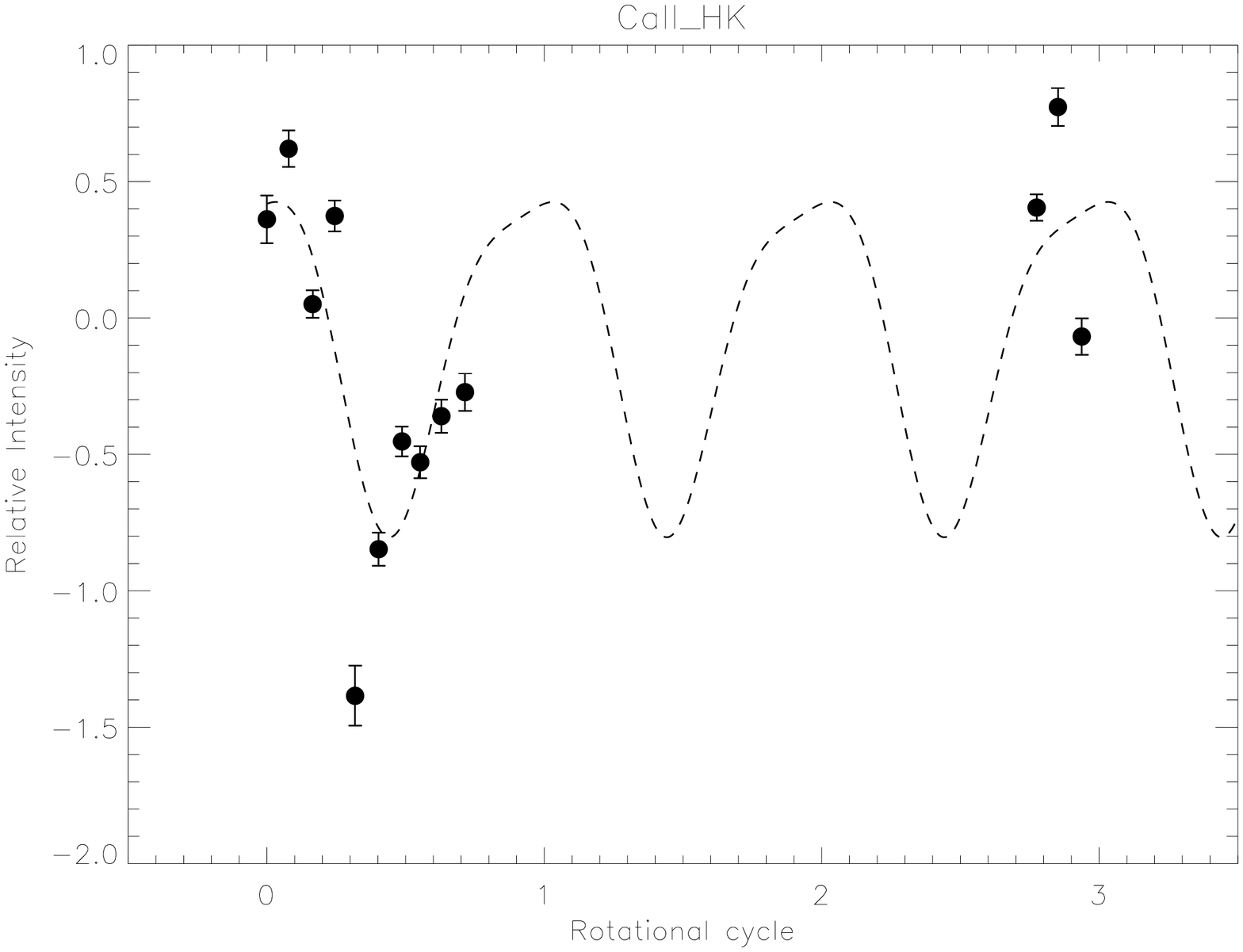}
\includegraphics[width=0.45\columnwidth]{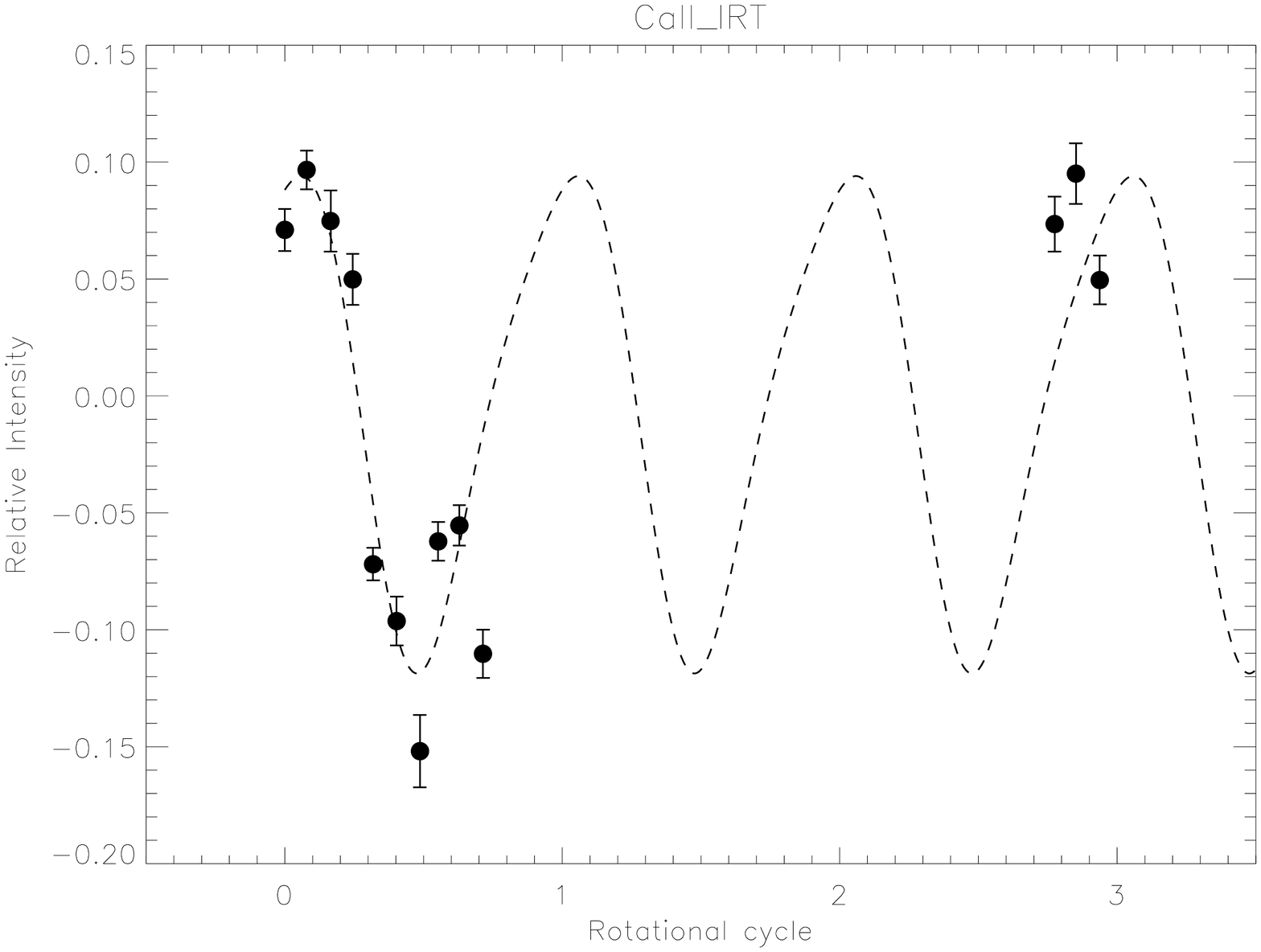}
\caption{The activity signals are modulated as a function of the rotation cycle of the star for H$\alpha$ (top left), H$\beta$ (top right), CaII H and K (bottom left) and CaII IRT (bottom right).  The dash lines show the best-fit model made of two sine waves at P$_{rot}$ and P$_{rot}$/2. Minimum of activity is observed when closed field lines are facing the observer.}
\label{modul}
\end{figure*}

We have then analyzed the activity proxies in the stellar spectra: CaII H and K lines in the 395 nm region, the CaII infrared triplet at about 850 nm, the H$\alpha$ and H$\beta$ lines at 656 and 486 nm (see Figure \ref{appenfig3b} for an example spectrum). Due to the magnitude and color of the star (V=11.72, B-V=1.15), the SNR per pixel in the range of the CaII H and K lines is low ($\sim$ 30). 
The Least Square Deconvolution profiles of all spectra were computed with a mask comprising only the relevant line(s); when several lines are included, a relative weight is applied to each line. Then, the average profile was subtracted to each individual profile, to get the residuals and enhance any detection of the variability. The residual profiles were fitted by a Gaussian and the area of this curve is our activity proxy; the values and their errors for H$_\alpha$, CaII H\&K and Ca IRT are given in Table \ref{log}, and all indices are plotted in Figure \ref{modul}. \\

This activity is modulated with time, and at first order, it varies with the stellar rotation period ($\sim$12.5 day), as shown on Figure \ref{modul}. Some data points, however, are not in good agreement with a simple sine wave. When a second sine wave at half the rotation period is added, then we obtain a better fit to the whole ESPaDOnS data set, as shown in Figure \ref{modul}. The modulation at P$_{rot}$/2 has an amplitude 2 to 6 times smaller than the P$_{rot}$ variation depending on the activity proxy. Such behaviour with a double modulation is common to the activity proxy of solar-type stars \citep[e.g.,][]{boisse10}. It was not necessary here to add a third signal with a P$_{rot}$ /3 frequency, as was done in previous activity studies of Kepler-78 \citep{pepe,howard}, because our data set contains very few points. The improvement of the fit due to the second period is mostly seen on the H$\alpha$ residual signatures; for CaII residuals, the addition of the first harmonic does not change the fit significantly. When the period is let free to vary, we find an optimal value of 11.9$\pm$0.4 days with all activity proxies included. It  means that the main active regions are located at a latitude of 20$\pm$16$^\circ$. These values are in good agreement with both the values derived from the magnetic map (section 4.2) and the values expected from the extrapolation of the spot cycle found in the $Kepler$ lightcurve (section 2). Finally, it is observed that a maximum of activity is seen in all proxies at rotational phases 0.1 and 2.9. Reversely, the minimum observed in most proxies at phase 0.5. We would rather expect the minimum of chromospheric activity to correspond to a region where field lines are open, at phase $\sim$0.2 (Fig. \ref{extrapolation}). Poor temporal coverage, possibly coupled to intrinsic variability, could easily explain the discrepancy we outline between activity proxies and the field line configuration. With the limitations of the data (noise and limited time series) and the reservations on the extrapolation techniques \citep{jardine2002} (mainly, the position of the source surface, see \citet{reville}, and the potential field assumption in the presence of a toroidal component), there is no cause for alarm about this apparent discrepancy. More data, densely sampled over 3 rotational periods, would be necessary to assess this point in detail, for example by testing different source surface locations or different extrapolation techniques, or using activity proxies and polarized profiles conjointly to reconstruct the surface topology. This has never been tested to our knowledge, and would benefit a stronger data set as a testbench. The possibility may still exist that the surface topology reconstruction described in section \ref{reszdi} is inaccurate, but it corresponds to the best we can infer from our limited set of data. The extrapolation map should be read with reservations in that context; it mainly illustrates the alternance of open and closed lines in the location of the planet orbit.
 
\begin{figure*}
\centering
\includegraphics[width=0.45\columnwidth]{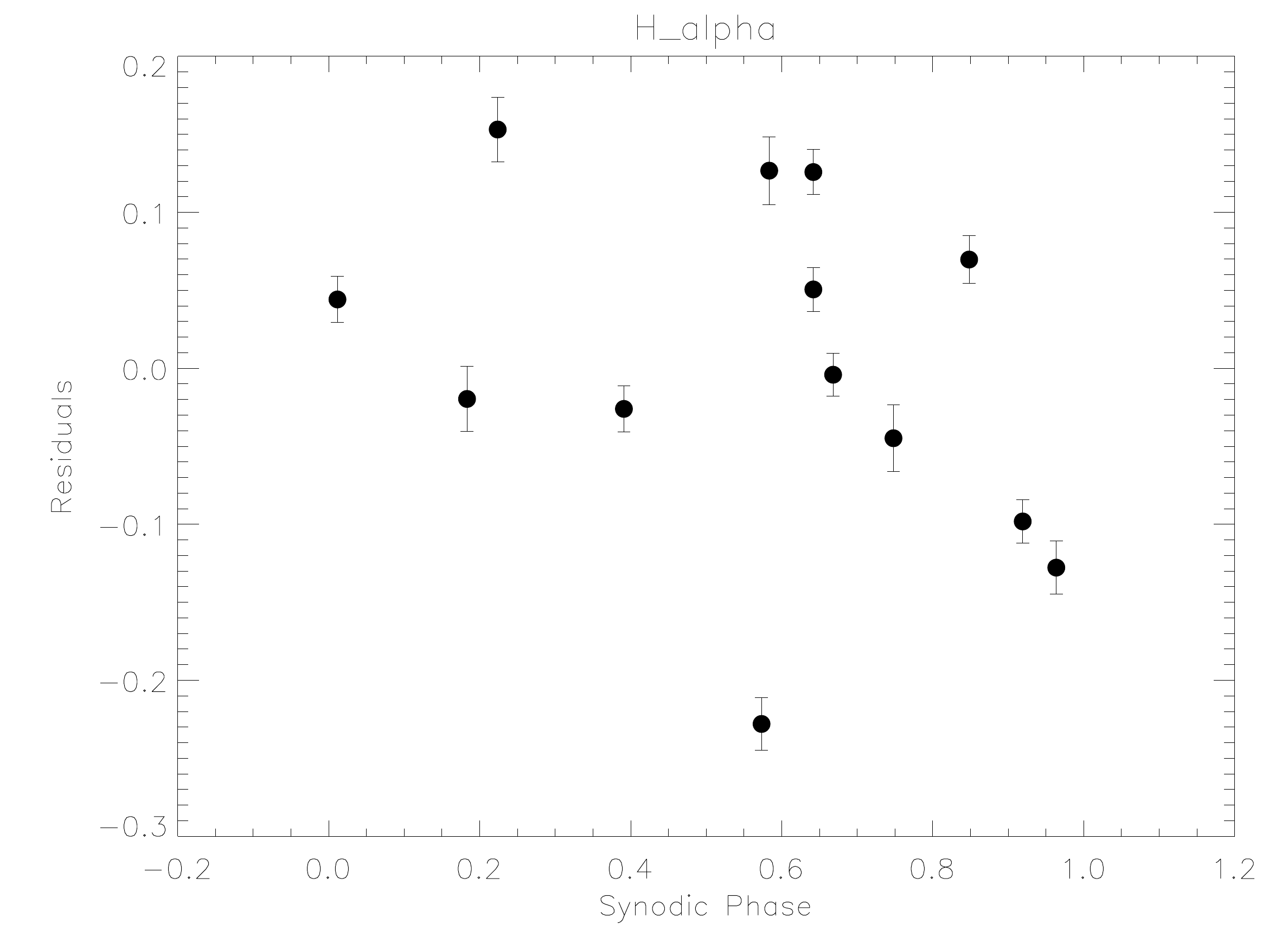}
\includegraphics[width=0.45\columnwidth]{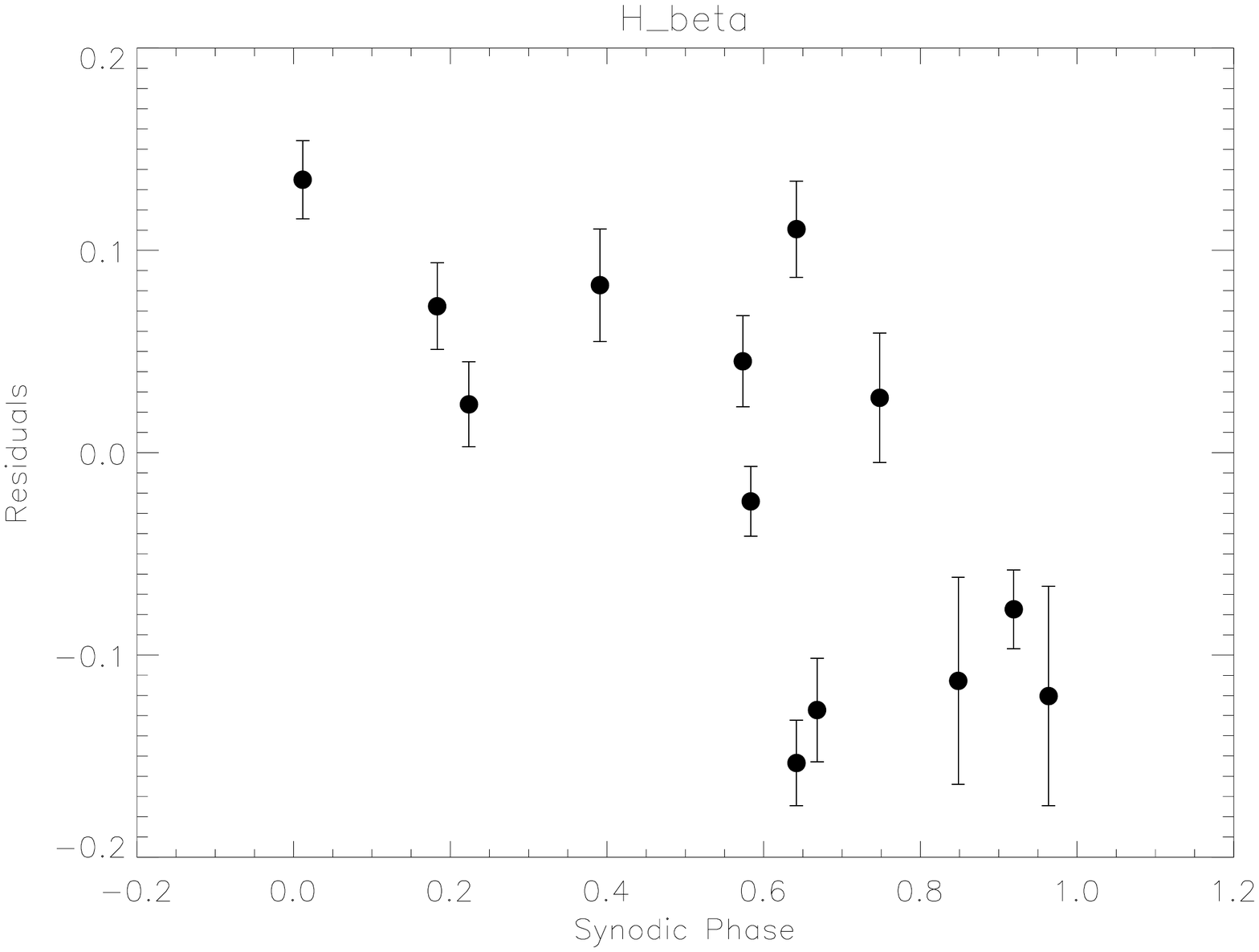}
\includegraphics[width=0.45\columnwidth]{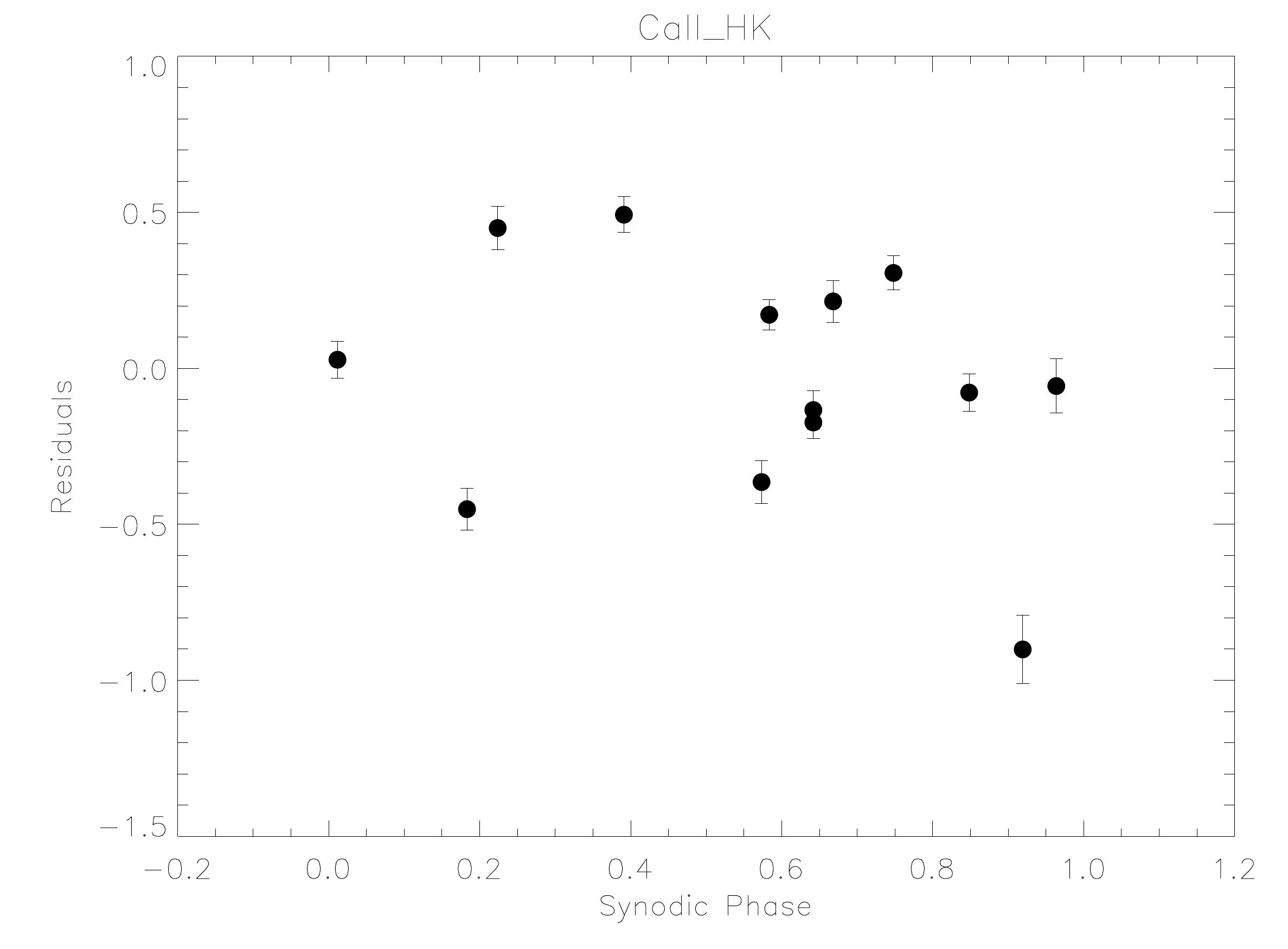}
\includegraphics[width=0.45\columnwidth]{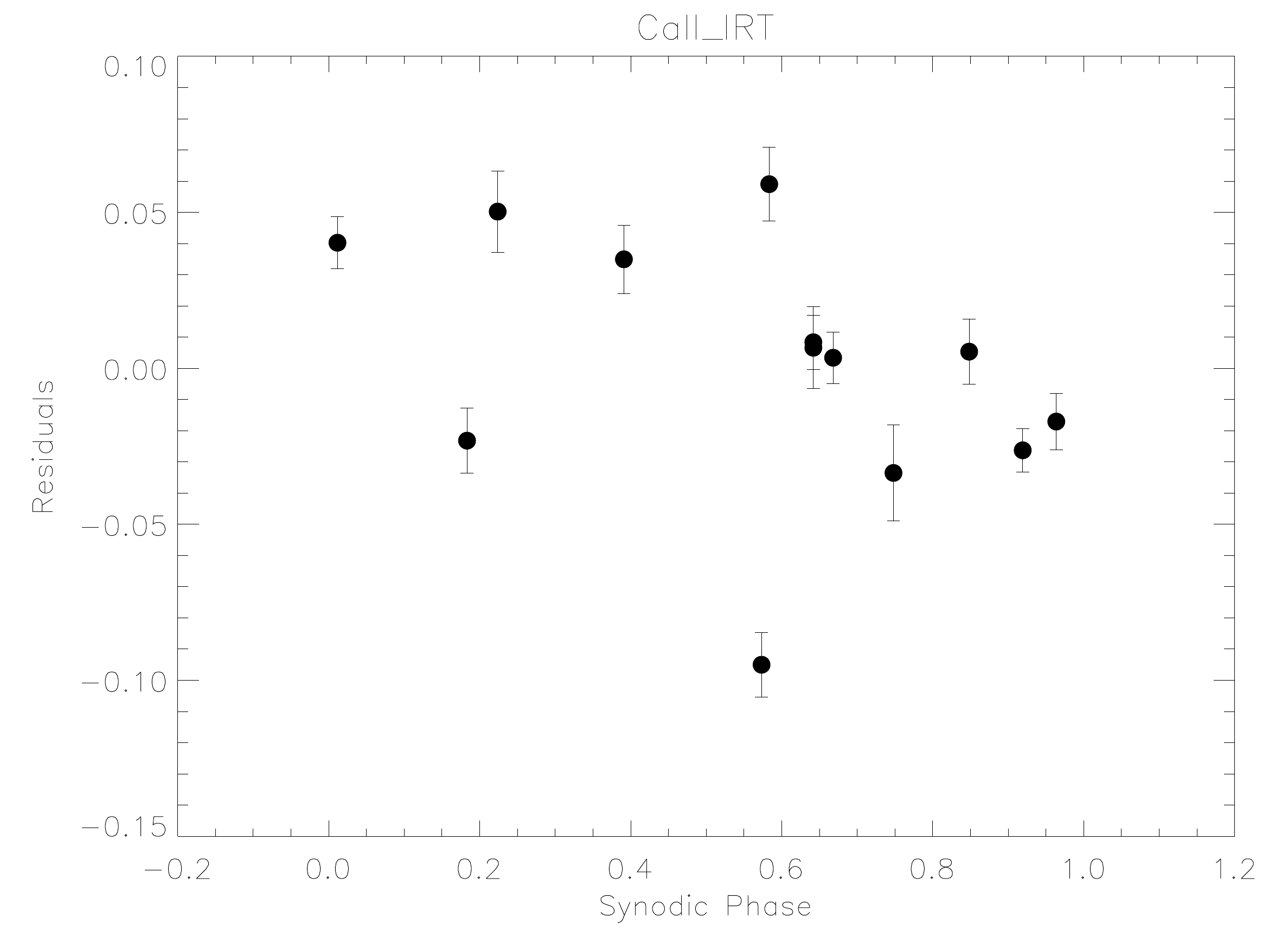}
\caption{The residuals of the CaII IR signal after the models shown on Figure \ref{modul} are removed, as a function of the synodic phase for various activity tracers: H$\alpha$ (top left), H$\beta$ (top right), CaII H and K (bottom left) and CaII IRT (bottom right).}
\label{resphsyn}
\end{figure*}

We also examined the activity signal in the light of the model of \citet{lainelin} that predicts the existence of intermittent hot spots on the stellar surface induced by the interactions with the close-in planet, and phased with the orbital or synodic period rather than the stellar rotational period. Figure \ref{resphsyn} shows the residual activity signal, once the rotational variability is removed, in phase with the synodic period, with ephemeris from \citet{sanchis13} and phase 0 at transit configuration (T$_{tr}$=245,453.95995). There is no hint for enhanced stellar variability at any synodic phase. The residual scatter is due either to some non-periodic intrinsic stellar variability, or an under-estimation of the errors. We find a very similar result when the freely-adjusted value of 11.9 days is used as rotational period.\\

\subsection{Looking for reflected light and exospheric signature from the planet}

While the stellar radial velocity variation induced by the planet is only $\sim$1.8 m/s \citep{howard, pepe}, the radial velocity semi-amplitude of the planet along its orbit is $\sim$ 250 km/s. This wide variation would put an hypothetical peak of the planet in the intensity profile far enough from the stellar profile to be looked at. We therefore conducted a search for an attenuated secondary peak in the stellar profile, using the same mask as for the star (5000K ATLAS9 model) since it is expected that the planet surface mainly emits by reflection at such a short distance. The search was done in a velocity range wide enough to include the putative planet peak, from -350 to +350 km/s. Note that the modulation of this signal is clearly seen on the $Kepler$ lightcurve, using the combination of about 3500 planetary orbits \citep{sanchis13}, with a maximum level at 6.10$^{-6}$ of the star. For the related spectroscopic signature, however, we expect an unfavorably large rotation effect: the stellar flux is seen by the planet at the synodic period of 0.365 day which corresponds to a velocity of 110 km/s, a factor ten broadening compared to the stellar intensity profile. The amplitude of the reflected light peak is thus reduced by a similar amount. Due to the duration of a polarimetric sequence, each spectrum also covers 20\% of the planet orbit and so it is expected that any signal of reflected light will be further smeared out, especially in the spectra where the planet radial velocity changes quickly during this phase span.

\begin{figure}
\centering
\includegraphics[width=0.5\columnwidth]{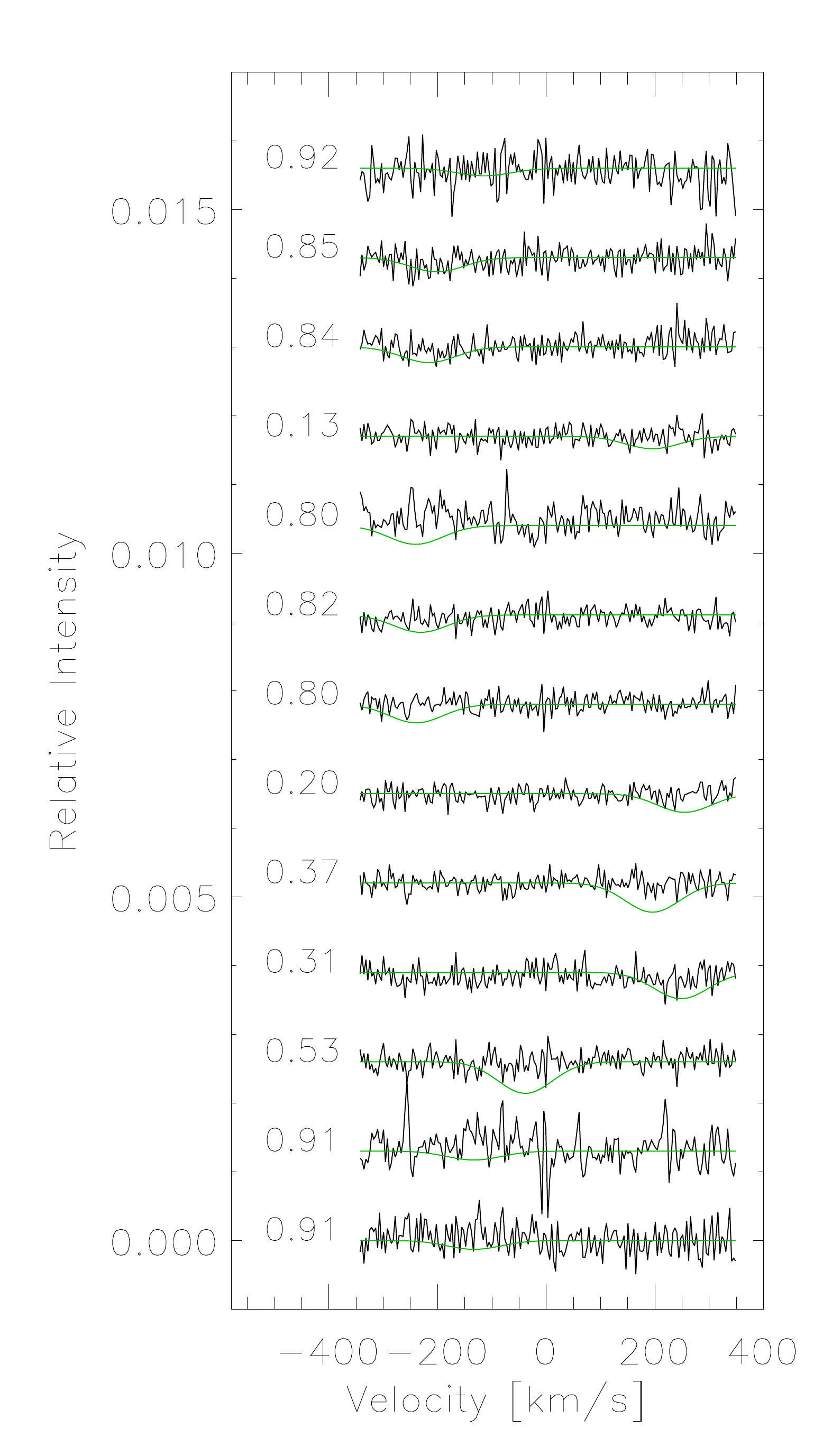}
\caption{The residual profiles of the intensity spectrum are shown, after the mean profile has been subtracted. The expected velocity of the planet is shown with superimposed green profiles where the amplitude has been amplified by a factor 1000. No signal is seen on this series of profiles. The orbital phase is given on the left of each spectrum, using the ephemeris of \citet{sanchis13} (phase equals 0 at transit times).}
\label{planetRV}
\end{figure}

The residual profiles were computed by subtracting a linear combination of the mean profile and its first derivative. Taking into account the first derivative allowed to cancel out the signal near the systemic velocity -3 km/s due to the instrumental instability and varying activity in the stellar peak. The $rms$ in the residual profiles ranges from 1.2 to 2.5 10$^{-4}$. The residual spectra are displayed in Figure \ref{planetRV}. 

Then we applied a matched filter analysis to the residuals to enhance the signal-to-noise ratio. We assumed that the planet signal due to reflected light has a known amplitude, width and position on each spectrum. By summing all pixels in the expected planetary profile and all epochs we expect to gain a factor $\sqrt{70\times15}$ if the noise follows Poisson statistics -- where 70 is the number of pixels in the planetary profile and 15 is the total number of spectra when all 2014 and 2015 data are included.  We thus expect to gain a factor 32 at most on the detection of the planetary signal reflected from the star; red noise is also present on the residual spectra, so we are not in a situation of Poisson noise statistics and the gain will be lower.

For each residual spectrum, we multiplied the spectrum by the expected signal in the form of a Gaussian function extended over 70 pixels. We modulated the amplitude of the reflected light with the orbital phase, with a maximum of 6.10$^{-6}$ times the stellar profile.  The expected reflected light peak is shown on the residual spectra in Figure \ref{planetRV}, amplified by a factor of 1000 for visibility.

\begin{figure}
\centering
\includegraphics[width=0.5\columnwidth]{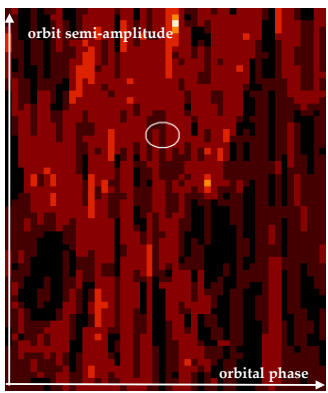}
\caption{Noise map of reflected light search as a function of orbital phase and semi-amplitude of the planet. The expected position of the planet is shown with the white circle. The map shows only noise.}
\label{matched}
\end{figure}

We varied the position of this Gaussian in radial-velocity semi-amplitude and phase of the planet orbit. 
Then we calculated the integrated value of all filtered spectra for a given pair of amplitude and phase, normalized by the square root of the number of pixels in the profile. The map of these values are plotted in Figure \ref{matched} as a function of phase (in abscissa) and amplitude (in ordinate). 
The reference time varies over the full orbit with step of 10 min and the semi-amplitude varies from 100 to 340km/s with steps of 4 km/s. The expected position of the planet is shown by the superimposed circle in the map, assuming the transit ephemeris from \citep{sanchis13} and the planet mass of \cite{pepe}. The signal is not enhanced at this location. The noise in the map has a standard deviation of 2.5 10$^{-5}$ and the signal at the expected planet location is at 0.9$\sigma$ level (2.22 10$^{-5}$). Hence, despite of a significant gain in sensitivity compared to the individual spectra (factor $\sim$10), we do not detect the reflected light emitted by the planet modulated by the planet motion.




Finally, since exospheric evaporation is also predicted for this kind of close-in system \citep[e.g.,][and references therein]{raymond08,ehrenreich}, we have looked for any signature in the spectra that could be due to photo-evaporated ions. We built an "exospheric mask" for profile calculations, made of all species expected to be present in the exosphere of a strongly irradiated telluric planet, as inspired from the work by \citet{guenther} on CoRoT-7 b: Sulfur II and III, CaII, NaI.  Again, no signature was detected at the planet velocity. The $rms$ in the residual  profiles are 15 times larger than with the full stellar mask, and ranges from 1.8 to 3.1 10$^{-3}$. In the theoretical expectation of \citet{ito15}, the secondary eclipse depth does not exceed a value of 10$^{-5}$ in the visible domain (the situation being more favorable in the infrared domain). Previous observations by \citet{guenther} had given upper limit contrasts of $\sim$ 10$^{-5}$ on CoRoT-7 b with VLT/UVES. In comparison, our data set is not of sufficient SNR for further investigations, except if many more predicted features could be added in the "exospheric mask". 


\section{Summary and discussion}

With the objective of characterizing the properties of Kepler-78's magnetic field, it was necessary to revisit the photometric periodicity of the $Kepler$ lightcurve, which both contains a wealth of information on the stellar activity and the signature of the ultra-short period planet Kepler-78 b \cite{sanchis13}. It was shown that the evolution in time of the apparent period of rotation which dominates the rapidly evolving spot signal shows a smooth modulation with a period of approximately 1300 days. We may interpret such behaviour as the activity cycle of the star produced by dynamo effects, with differential rotation and spots migrating towards lower latitudes under the action of dynamo processes and the resulting cyclically varying large-scale magnetic field. Since the $Kepler$ lightcurve covers a 1560 day period, it is yet a bit short to confirm any periodic behaviour at such scale; further photometric monitoring could confirm, or refute this interpretation. Also, ultimately, it is only the reversal of the large-scale magnetic field which would demonstrate the existence of an activity cycle similar to the solar cycle, for which further spectropolarimetric campaigns would be required on a few years timescale.
From this analysis, we find a differential rotation larger than the Sun, of at least $\sim$0.08 rad.d$^{-1}$. This is confirmed by the differential rotation rate  of 0.105$\pm$0.039 rad.d$^{-1}$ found by analyzing the spectropolarimetric observations with Zeeman Doppler Imaging, although this detection of differential rotation is just a hint and more spectropolarimetric data would be necessary to confirm it. The rotation properties and main active latitudes found in our ESPaDOnS data (magnetic analysis and chromospheric proxies) are in agreement with the expectation from extrapolating the $Kepler$ lightcurve activity cycle.

Such complex and high-quality lightcurve would benefit, however, from a more complete spot modeling as done for other $Kepler$ active planet hosts like Kepler-17 \citep{bonomolanza}. The similarity between Kepler-78 and Kepler-17 is important to note. Both stars are young GK dwarfs (Kepler-17 being both a bit more massive and older than Kepler-78) and rotate in about 12 days. Their $Kepler$ lightcurves behave similarly, although Kepler-78's amplitude is ten times lower than Kepler-17's; the differential rotation derived from spot modelling by \citet{bonomolanza} lies in the range 0.052-0.084 rad.d$^{-1}$ for Kepler-17, while it is $\sim$ 0.1 rad.d$^{-1}$ for Kepler-78. As a sanity check, we performed a similar analysis on the $Kepler$ light curve of Kepler-17: we calculated the highest power peak in the periodogram for each quarter where $Kepler$  observed this star (see Figure \ref{appenfig2} in Appendix). We find a value for this peak (the main rotational period) ranging from 11.4 to 12.9 days, which correspond to a differential rotation larger than 0.064 rad.d$^{-1}$. We also find no cyclic behaviour of this period over the 4 years of $Kepler$ data (see Figure \ref{appenfig2b}). The measured value of 0.064 rad.d$^{-1}$ is in good agreement with the range given by the spot modeling by \citet{bonomolanza}.  More young solar-type stars have differential rotations in the range 0.08-0.45 rad.d$^{-1}$ as shown by, e.g., \citet{frohlich} and \citet{marsden}. Also, the well-studied hot-Jupiter host HD 189733 has a differential rotation of 0.146$\pm$0.049 rad.d$^{-1}$ \citep{fares10}, a mean rotation period of about 12.5 days, and no reported cycle over a time span of a few years.\\

The lap time derived from our estimate of the differential rotation is 2$\times \pi / d\Omega \sim $ 60 days; this corresponds to the time needed for the pole to make a revolution less than the equator. On the other hand, \citet{pepe} report an e-folding timescale of about 50 days from the analysis of the $Kepler$ lightcurve, that these authors attribute to the typical lifetime of spots. The agreement between these two values would suggest that differential rotation could control the lifetime of spots -- at least for those which dominate the radial-velocity signal -- rather than normal or turbulent magnetic diffusivity \citep[e.g.][]{bradshaw}.


 
 Another interesting comparison can be done with $\epsilon$ Eri, a young K2 dwarf that has chromospheric activity with a  3 year period \citep{hatzes00} that is modulated by a longer 13 year period \citep{metcalfe}. The long-term evolution of the magnetic topology of this star, however, is complex and not cyclic over 7 years of monitoring \citep{jeffers}.  It should thus be repeated that the best indication of a dynamo cycle remains a clear inversion of polarity, a mechanism which has been witnessed in very few stars so far: several polarity reversals for $\tau$ Bootis \citep{donati08,fares09} and 61 Cyg A (Boro Saikia et al, in prep.) and one reversal for HD 190771 \citep{petit}.  To witness these polarity reversals, one must observe repeatedly the same stars, derive the topology of their surface over time, and search for a change of sign in the main polarity of a given magnetic hemisphere.\\

Cyclic activity is important for star-planet interactions. Previous studies have invoked intrinsic variations in the stellar extended magnetic envelopes at the origin of intermittent phenomena, as the enhancement of CaII emission modulated by the orbital period \citep{shkolnik} or changes in the upper atmosphere of a planet \citep{lecavelier}. It is then important to get a more global understanding of the stellar activity timescales, and of the amplitude of this variation. This task is, however, enormous due to the large quantity of data necessary to reconstruct the magnetic topology at a given epoch (typically 22 hours of observations on a 3.6m telescope for a rather faint star like Kepler-78). Ideally, each search for a specific star-planet interactions signature should be joined by a contemporaneous campaign in spectropolarimetry to constrain the input stellar conditions. Future transiting planetary systems to be discovered by TESS and PLATO will be hosted by brighter targets for which spectropolarimetric observations will be easier to conduct repeatedly.\\

The magnetic topology of Kepler-78 is comparable to other stars of similar mass and rotation rate. The mean strength of the magnetic field is 16 G at the time of observations, with 60\% of the magnetic energy being in the poloidal component. The planet crosses alternately closed field lines and open field lines along its orbit at 3 stellar radii. The strength of the magnetic field as viewed by the planet varies along the orbit with an amplitude which depends of the assumptions made for the location of the source surface in extrapolating the surface field. We would expected that the minimum of chromospheric activity corresponds to phases where the line of sight crosses the wide area with open field lines (coronal hole), but rather, this minimum is shifted by about 0.2 in phase. This shift could be a consequence of the strong differential rotation inducing shearing in the chromosphere, but at the moment it is not understood and would require additional monitoring.

We used a matched filter to search for the signature of the stellar light reflected by the planet in the spectra, and reached a detection limit of 2.5 10$^{-5}$. The apparent broadening of the stellar peak reflected by the planet is a severe limitation of the method, compared to photometry. This detection level is expectedly not sufficient to find the signal and we report a negative result still in agreement with the previous detection by $Kepler$ \citep{sanchis13}. The observation strategy was also not optimized for this search, and many spectra do not correspond to an orbital phase where the planet is bright.

Finally, this study was conducted in the context of star-planet interactions, and the recent prediction that hot spots could be induced at the stellar surface due to Ohmic dissipation transmitted in the footprint of the planet in ultra-short orbit \citep{lainelin}. 
The measured value of the mean magnetic field implies that the unipolar induction is an important process for the planet-star interaction.  The permeation of the time dependent component of the field into the planet's interior may also lead to episodic Ohmic dissipation.  These quantities combined with the upper limits on the planet's orbital decay rate and intrinsic intensity provide useful constraints on the theoretical inference of the planet's conductivity to be $\sim$ 0.01-0.1 S/m which is comparable to that of partially molten rock (Laine \& Lin in preparation).

Our ESPaDOnS observations did not reveal hot spots in the activity tracers at the orbital nor synodic phases. This non-detection should not, however, prevent any further theoretical work in this direction. 
Kepler-78 is intrinsically an active star, its activity signal is primarily related to the stellar rotation cycle, and any residual to the main spot rotation modulation could still be due to flares or other intermittent activity demonstration at the star surface, that are not uncommon. Potential planet-induced hot spots even at a level of $\sim$ 10\% may remain hidden by the intrinsic variability of the stellar surface. Finally, the timescales of apparition of the planet-induced hot spots, as well as their lifetime, could well be much shorter than the individual exposure time of our data. Due to the relatively faint magnitude of Kepler-78, we had to collect photons for about 30 minutes per exposure, or 2 hours in total for a nightly visit. If planet-induced hot spots are of the same order of duration than the planet transit (48 min), then our observational sampling is not adequate. Short cadence H$\alpha$ photometric monitoring could be more appropriate, and, if combined with contemporaneous spectropolarimetry, could help distinguishing intrinsic stellar variations from planet induced activity. Future observational work in this direction could also focus on less intrinsically variable stars with ultra-short planets, although the stellar magnetic field would then be more difficult to constrain.\\

The detection of hot spots, with a luminosity higher than the stellar irradiation received by the planet, would have been an unequivocally signature of the unipolar induction process. However, the ESPaDOnS's quantitative upper limit places a useful constraint on the intensity of Ohmic dissipation at the footprint of the flux tube which connects Kepler-78 and its close-in planet.  This constraint is consistent with that inferred from the upper limit of the orbital decay rate for Kepler-78 b (Laine and Lin, in prep.).

Non detection of stellar spots which co-rotate with the planet's orbit at intensity level substantially below the stellar irradiation on the planet would indicate that unipolar induction between Kepler-78 b and its host star may be interrupted by field reconnection or there is insufficient time for the Alfven waves to complete the circuit.  These theoretical implications will be presented elsewhere.

\appendix
\section{Quarterly periodograms from the $Kepler$ lightcurves}
\begin{figure}
\includegraphics[width=0.5\columnwidth]{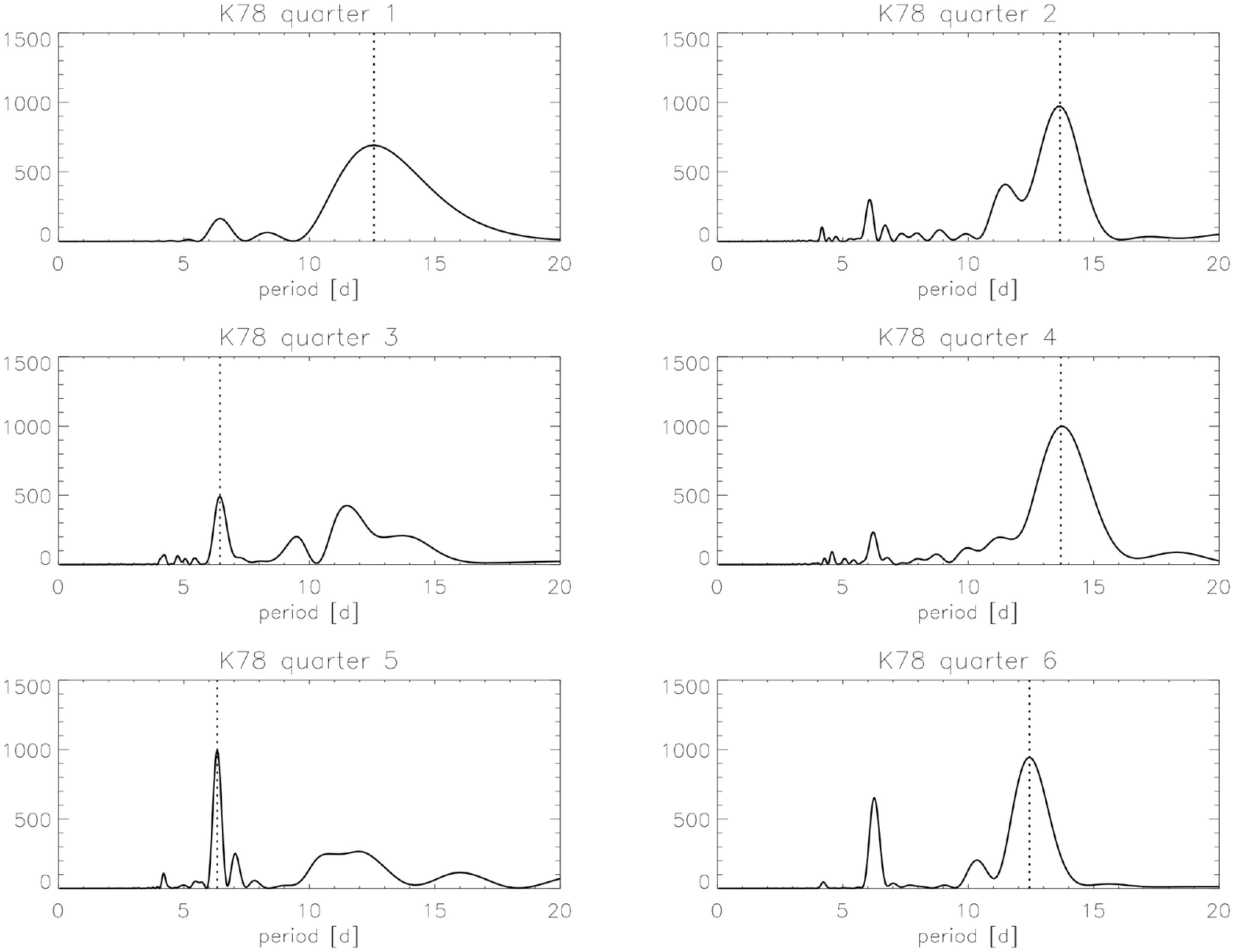}
\includegraphics[width=0.5\columnwidth]{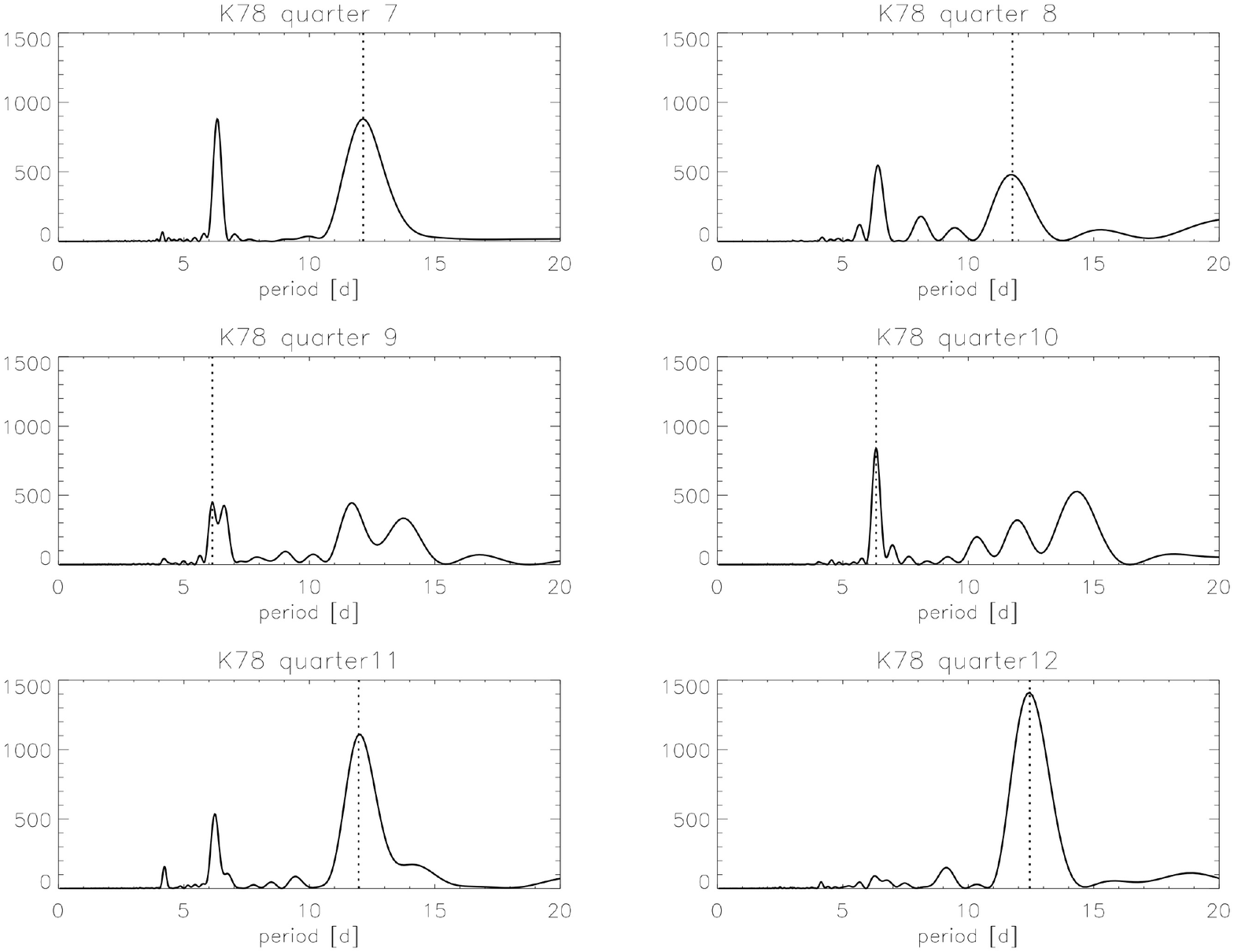}
\includegraphics[width=0.5\columnwidth]{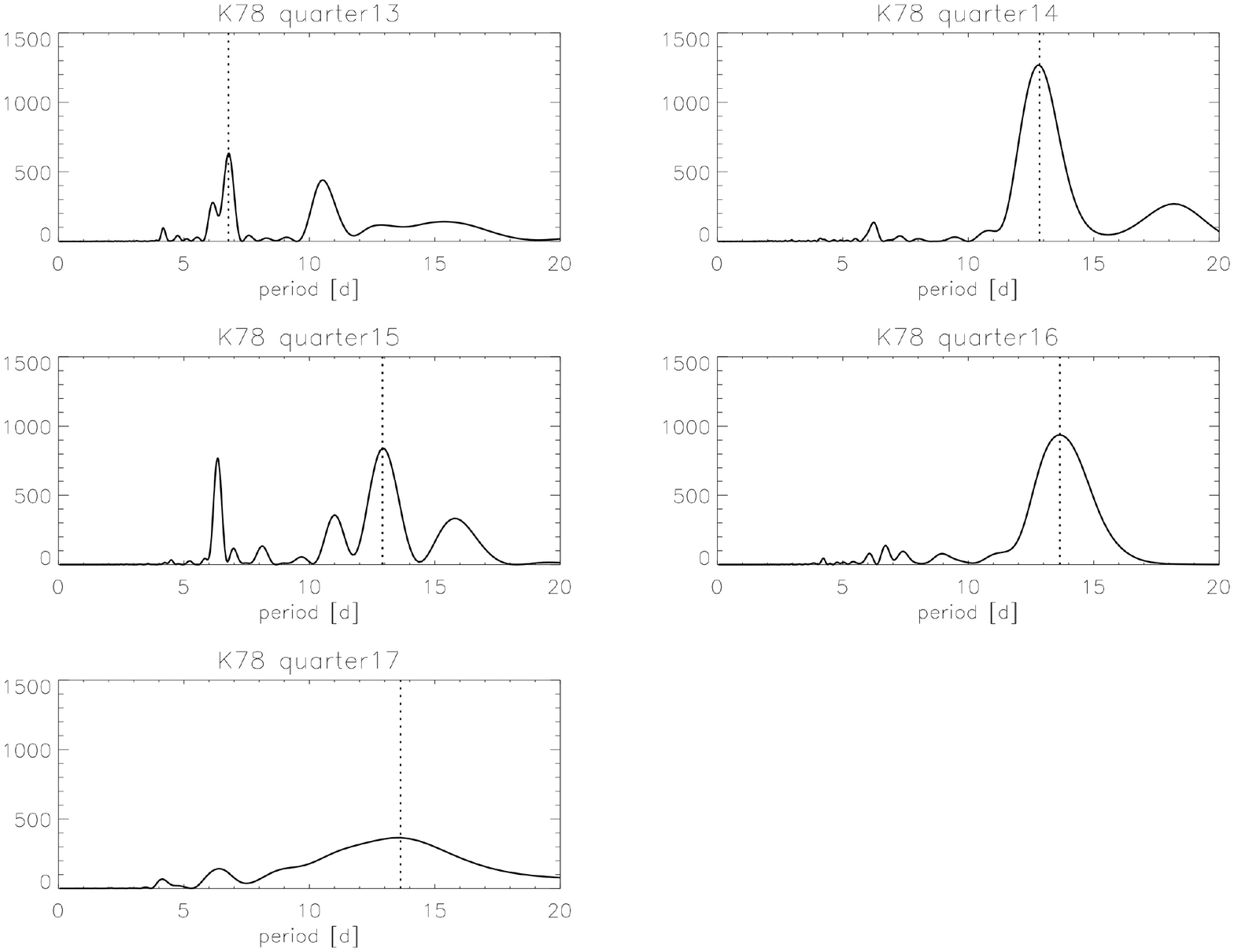}
\caption{Periodograms of the $Kepler$ light curve of Kepler-78 for each quarter. Each vertical line shows the main peak obtained from fitting a Gaussian to the peak, and reported in Figure 1.}
\label{appenfig1}
\end{figure}

\begin{figure}
\includegraphics[width=0.5\columnwidth]{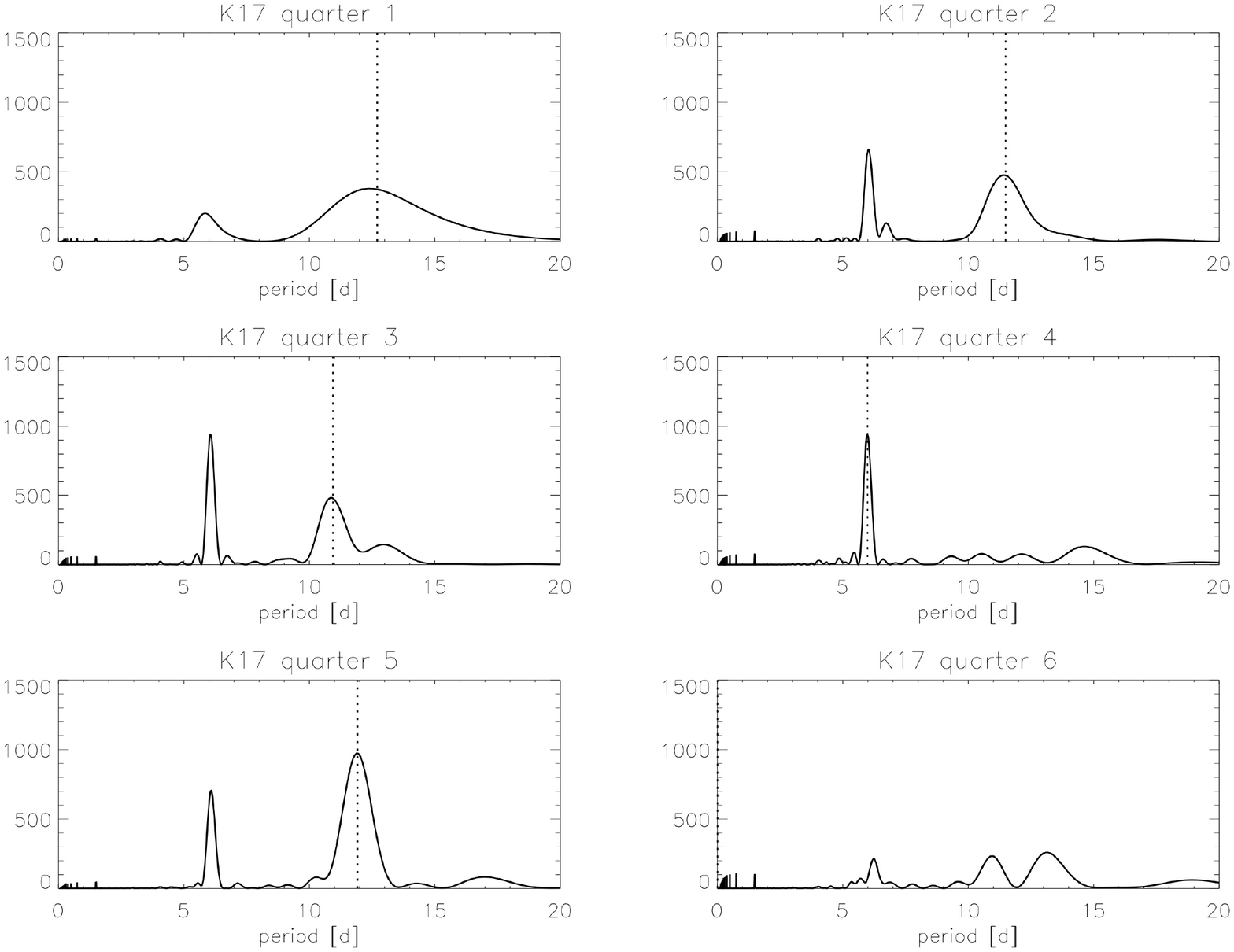}
\includegraphics[width=0.5\columnwidth]{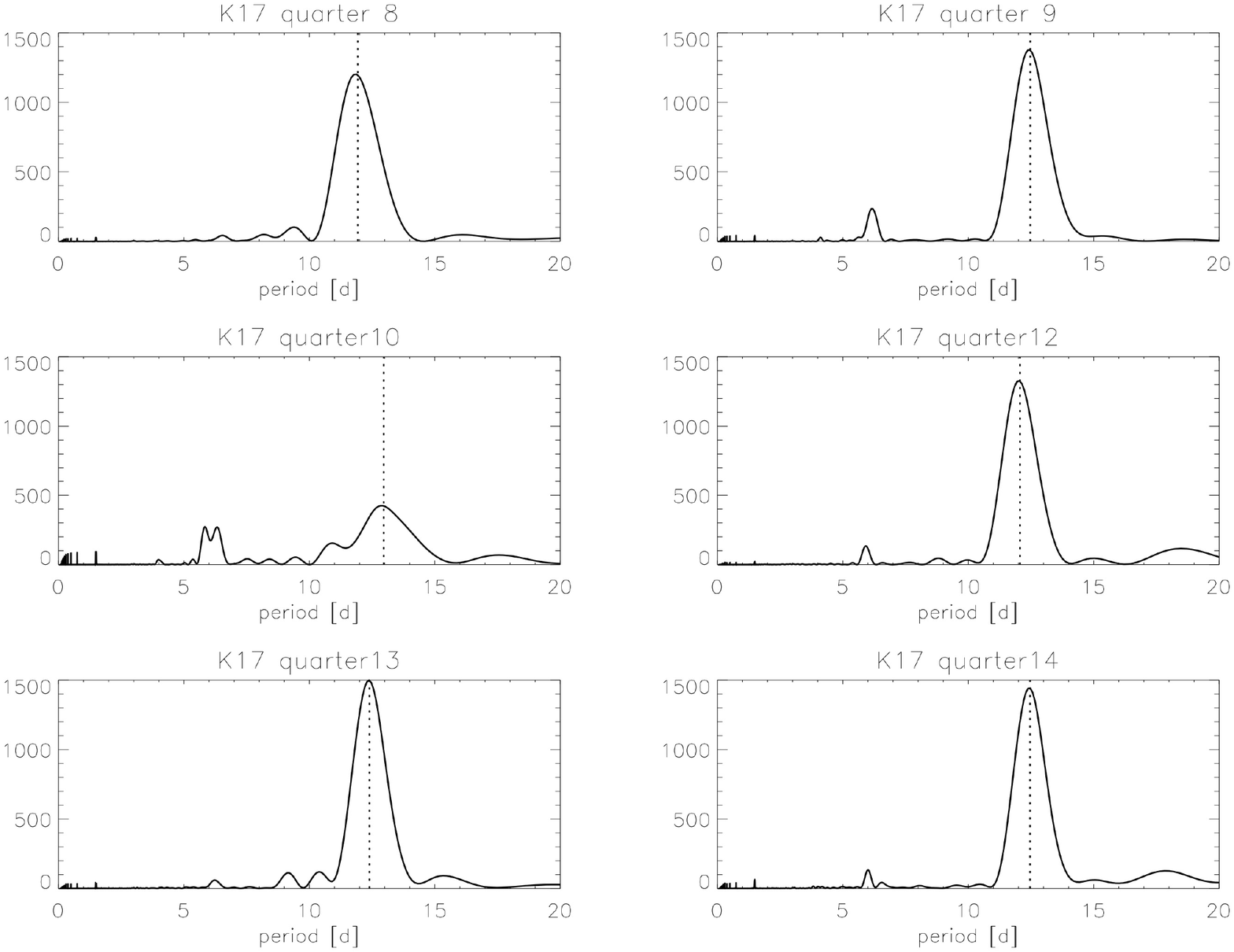}
\includegraphics[width=0.5\columnwidth]{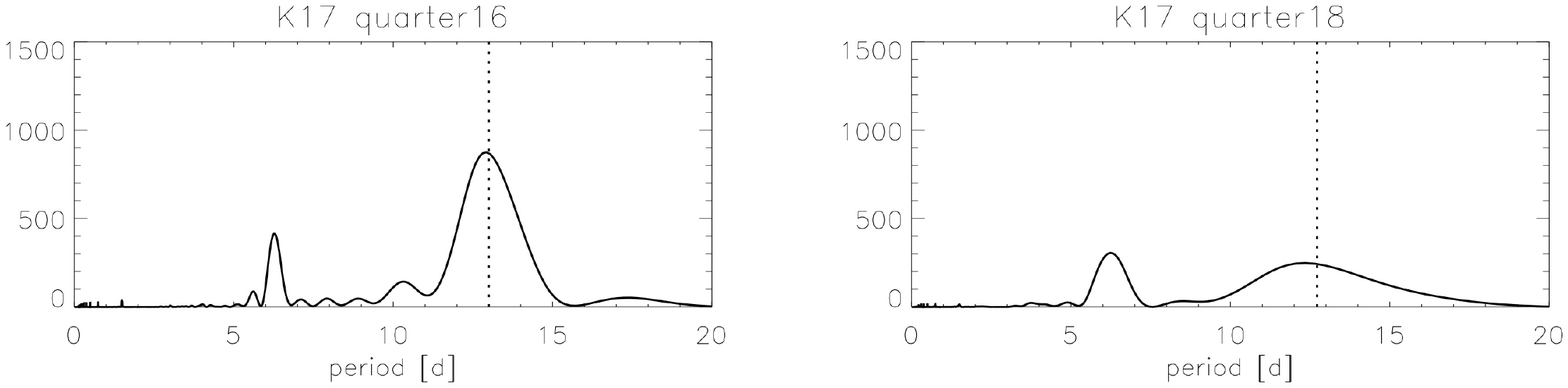}
\caption{Periodograms of the $Kepler$ light curve of Kepler-17 for each quarter when the star was observed.  }
\label{appenfig2}
\end{figure}

\begin{figure}
\includegraphics[width=1\columnwidth]{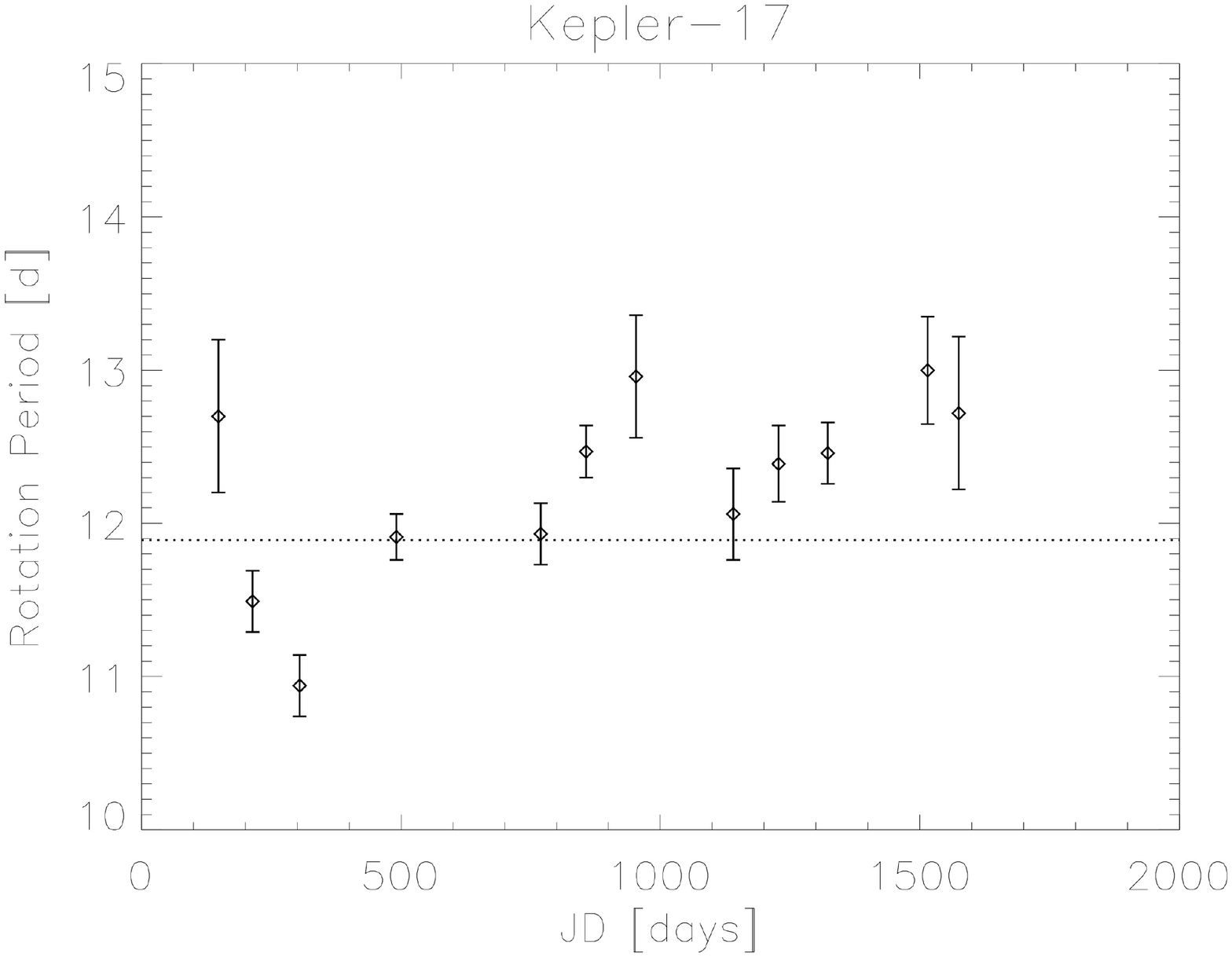}
\caption{The main rotation period of Kepler-17 as a function of time during the full duration of the $Kepler$ mission. }
\label{appenfig2b}
\end{figure}

Figure \ref{appenfig1} shows the periodograms obtained on $Kepler$ data of Kepler-78 for each quarter. These periodograms were used to measure the apparent rotational period plotted in Figure \ref{rot} and its variation in time. \\
As mentionned in the discussion, the same method was then applied to another host star with strong rotational modulation, Kepler-17. The periodograms for each $Kepler$ quarter are shown on Figure \ref{appenfig2}. Figure \ref{appenfig2b} shows the behaviour in time of the derived rotation period. Contrarily to Kepler-78, there is no apparent periodic modulation of the rotation period of Kepler-17 over 4 years, which may indicate a longer activity cycle than for Kepler-78.

\section{Example spectra of Kepler-78}
\begin{figure}
\includegraphics[width=1\columnwidth]{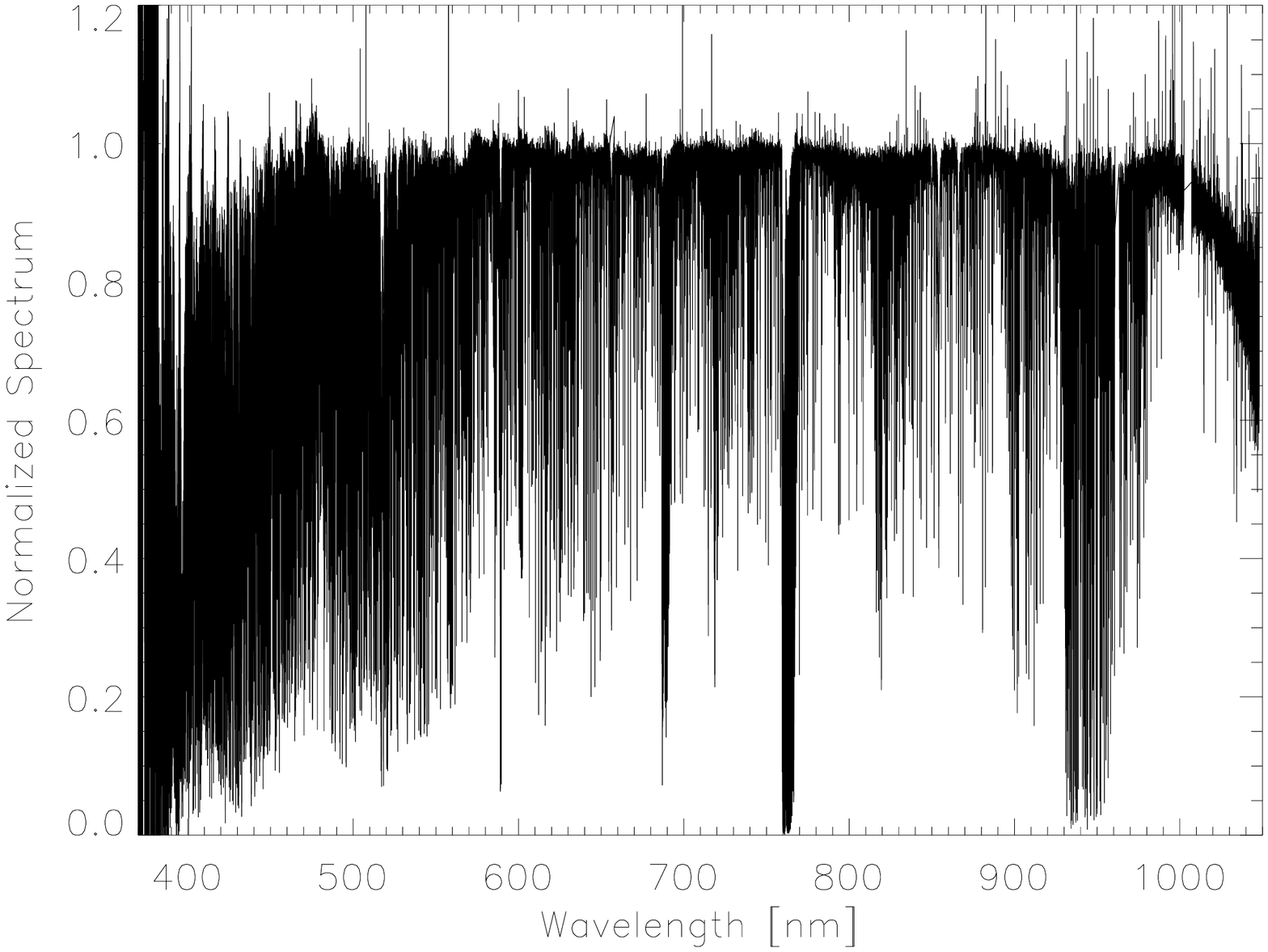}
\caption{An example spectrum of Kepler-78, obtained on September 7, 2014.}
\label{appenfig3}
\end{figure}

\begin{figure*}
\centering
\includegraphics[width=1\columnwidth]{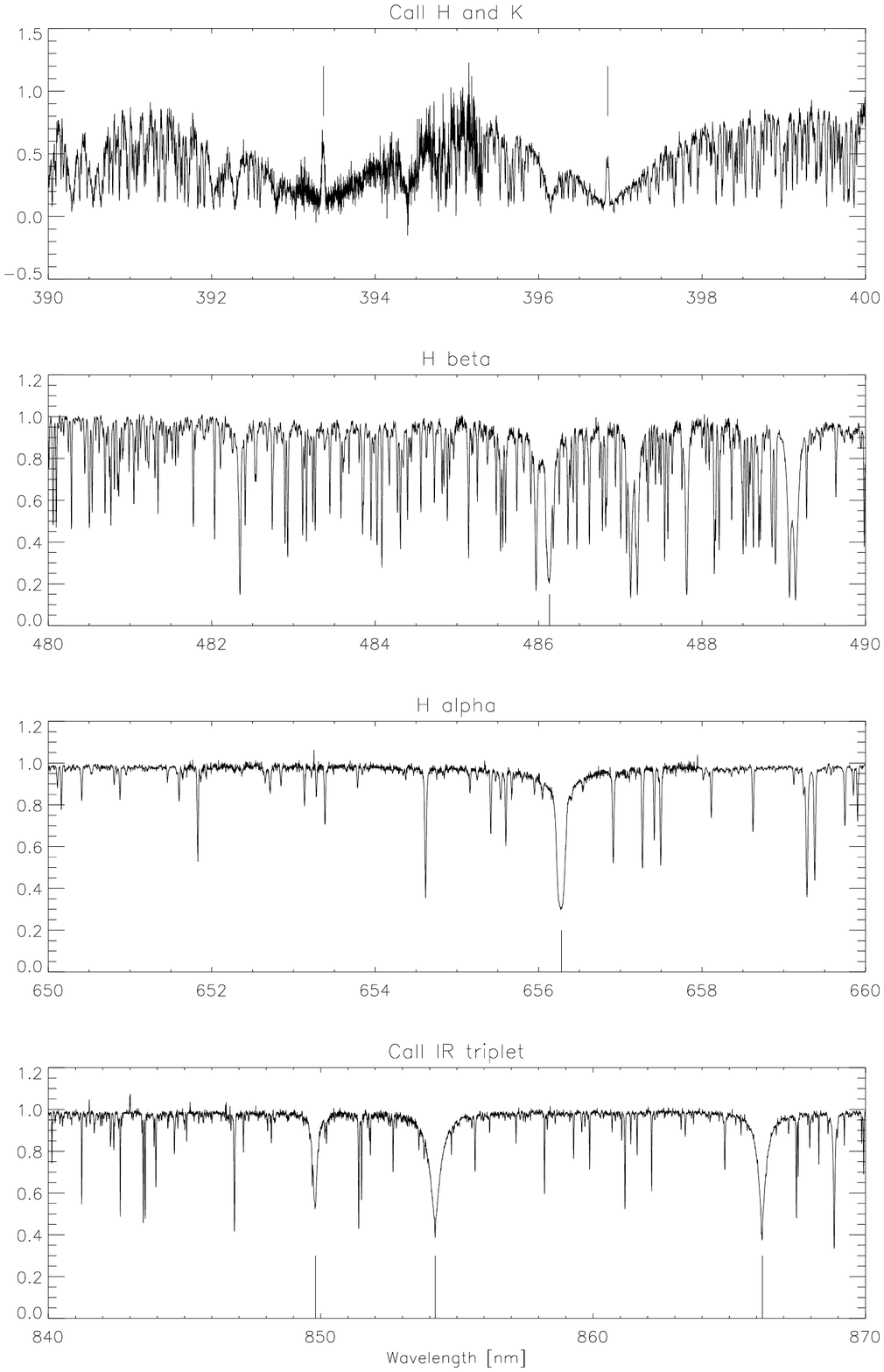}
\caption{One example spectrum is shown, around the four activity proxy lines which time evolution is described in the text: from top to bottom, CaII H and K lines, H$_\beta$, H$_\alpha$ and CaII infrared triplet. Vertical ticks show the position of the lines.}
\label{appenfig3b}
\end{figure*}

Figure \ref{appenfig3} shows an example ESPaDOnS spectrum of Kepler-78 (07Sept2015), as extracted and normalized by Libre-Esprit, in the full spectral domain. Figure \ref{appenfig3b} shows the same spectrum zoomed around the activity lines of CaII and HI which evolution is discussed in the text.

\small

\label{lastpage}
\end{document}